\documentclass{article}
\usepackage{amsmath}
\usepackage{amsfonts}
\usepackage{amsthm}
\usepackage{amssymb}

\newtheorem{theorem}{Theorem}[section]

\newtheorem{lemma}[theorem]{Lemma}
\newtheorem{corollary}[theorem]{Corollary}

\input{tcilatex}

\begin{document}

\title{Bounds on the Pure Point Spectrum of Lattice Schr{\"o}dinger Operators%
}
\author{V. Bach, W. de Siqueira Pedra, S. N. Lakaev}
\date{18-September-2017}
\maketitle

\begin{abstract}
In dimension $d\geq 3$, a variational principle for the size of the
pure point spectrum of (discrete) Schr{\"{o}}dinger operators $H(\mathfrak{e}%
,V)$ on the hypercubic lattice $\mathbb{Z}^{d}$, with dispersion relation $%
\mathfrak{e}$ and potential $V$, is established. The dispersion relation $%
\mathfrak{e}$ is assumed to be a Morse function and the potential $V(x)$ to
decay faster than $|x|^{-2(d+3)}$, but not necessarily to be of definite
sign. Our estimate on the size of the pure-point spectrum yields the absence
of embedded and threshold eigenvalues of $H(\mathfrak{e},V)$ for a class ot potentials of this kind. 
The proof of the variational
   principle is based on a limiting absorption principle combined with
   a positive commutator (Mourre) estimate, and a Virial theorem. A
   further observation of crucial importance for our argument is that,
   for any selfadjoint operator $B$ and positive number $\lambda >0$,
   the number of negative eigenvalues of $\lambda B$ is independent of
   $\lambda$. \\[0.9ex]
\textit{Key words.} discrete Schr{\"{o}}dinger operators; embedded
eigenvalues.\\[0.5ex]
\textit{2010 Mathematics Subject Classification.} 46N50; 81Q10.
\end{abstract}

\section{Introduction}

Let $\Gamma \doteq \mathbb{Z}^{d}$ be the $d$-dimensional hypercubic
lattice. Given a bounded potential $V \in \ell^{\infty}(\Gamma; \mathbb{R})$%
, the discrete Schr\"{o}dinger operator corresponding to $V$ is 
\begin{equation*}
-\Delta _{\Gamma }+V\text{ }, 
\end{equation*}
where $V$ acts as a multiplication operator and $\Delta _{\Gamma }$ is the
discrete Laplacian defined by 
\begin{equation*}
\lbrack \Delta _{\Gamma }{\psi }](x)\doteq \sum_{|v|=1}\left\{ {\psi }(x+v)-{%
\psi }(x)\right\} \text{ }. 
\end{equation*}
More generally, we assume to be given a function $\mathfrak{e}\in
C^{2}(\Gamma ^{\ast };\mathbb{R})$ on the $d$-dimensional torus (Brillouin
zone) $\Gamma ^{\ast }\doteq \left( \mathbb{R}/2\pi  \mathbb{Z}\right)
^{d}\equiv \lbrack -\pi ,\pi )^{d}$, the dual group of $\Gamma $. We refer
to $\mathfrak{e}$ as a \textit{dispersion relation} or simply a \textit{%
dispersion}. We then consider the self-adjoint operator 
\begin{equation*}
H(\mathfrak{e},V)\ \doteq \ h(\mathfrak{e})+V\text{ }, 
\end{equation*}%
on $\ell^{2}(\Gamma )$, where $h(\mathfrak{e})\in \mathcal{B}%
[\ell^{2}(\Gamma )]$ is the hopping matrix (convolution operator)
corresponding to the dispersion relation $\mathfrak{e}$, i.e., 
\begin{equation*}
\left[ \mathcal{F}\left( ^{\ast }h(\mathfrak{e}){\psi }\right) \right](p) \
= \ \mathfrak{e}(p)\,[\mathcal{F}^{\ast }({\psi })](p)\text{ }, 
\end{equation*}
for all ${\psi }\in \ell ^{2}(\Gamma )$. Here, 
\begin{equation*}
\mathcal{F}^{\ast }:\ \ell ^{2}(\Gamma )\rightarrow L^{2}(\Gamma ^{\ast
}),\quad \lbrack \mathcal{F}^{\ast }({\psi })](p)\ \doteq \ \sum_{x\in
\Gamma }\mathrm{e}^{-i\langle p,x\rangle }{\psi }(x)\text{ }, 
\end{equation*}
is the usual discrete Fourier transformation with inverse 
\begin{equation*}
\mathcal{F}:\ L^{2}(\Gamma ^{\ast })\rightarrow \ell ^{2}(\Gamma ),\quad
\lbrack \mathcal{F}(\psi )](x)\ \doteq \ \int_{\Gamma ^{\ast }}\mathrm{e}%
^{i\langle p,x\rangle }\psi (p)\,\mathrm{d}\mu ^{\ast }(p)\text{ }, 
\end{equation*}%
where $\mu ^{\ast }$ is the (normalized) Haar measure on the torus, $\mathrm{%
d}\mu ^{\ast }(p)=\frac{\mathrm{d}^{d}p}{(2\pi )^{d}}$. Put differently, $h(%
\mathfrak{e})=\mathcal{F}\mathfrak{e}\mathcal{F}^{\ast }$ is the Fourier
multiplier corresponding to $\mathfrak{e}$.

For each $x \in \Gamma $, let $\delta_{x} \in \ell ^{2}(\Gamma )$ be the
normalized vector 
\begin{equation*}
\delta_{x}(y)\doteq \delta _{x,y}\text{ }, 
\end{equation*}
where $\delta_{x,y}$ is the Kronecker delta. For a dispersion relation $%
\mathfrak{e}$ and a pair $(x,y)\in \Gamma^{2}$, define the hopping amplitude 
\begin{equation*}
h(\mathfrak{e})_{xy}\doteq \left\langle \delta_{x}| \, h(\mathfrak{e})
\delta_{y} \right\rangle \text{ }. 
\end{equation*}
We say that $h(\mathfrak{e})$ has a \textit{finite range} if, for some $R <
\infty$ and all $(x,y)\in \Gamma ^{2}$, $h(\mathfrak{e})_{xy}=0$ when $|x-y|
\doteq \sqrt{(x_1-y_1)^2 + \ldots + (x_d-y_d)^2} >R$. The smallest number $R(%
\mathfrak{e}) \geq 0$ with this property is the \textit{range} of the
hopping matrix $h(\mathfrak{e})$. Equivalently, $h(\mathfrak{e})$ has a
finite range whenever $\mathfrak{e}$ is a trigonometric polynomial.

W.l.o.g., the minimum of $\mathfrak{e}$ is assumed to be $0$, so 
\begin{equation*}
\mathfrak{e}(\Gamma ^{\ast }) \ = \ [0,{\mathfrak{e}_{\mathrm{max}}}(%
\mathfrak{e})] \text{ }. 
\end{equation*}
We will further assume that the dispersion $\mathfrak{e}$ satisfies the
following condition: 
\begin{equation*}
(\mathbf{M}) \qquad \mathfrak{e}\text{ and }|\nabla \mathfrak{e}|^{2}\doteq
\sum_{k=1}^{d}\left\vert \partial _{p_{k}}\mathfrak{e}\right\vert ^{2}\text{
are Morse functions.} 
\end{equation*}
Clearly, the condition (\textbf{M}) is stable under small perturbations in
the $C^{3}$-sense, i.e., if $\left\Vert \mathfrak{e}^{\prime }-\mathfrak{e}%
\right\Vert _{C^{3}(\Gamma ^{\ast })}$ is sufficiently small and ${\mathfrak{%
e}}$ satisfies (\textbf{M})\textbf{, }then so does $\mathfrak{e}^{\prime }$.
Moreover, if the dispersion ${\mathfrak{e}}$ has a finite range then so do $%
|\nabla \mathfrak{e}|^{2}$, with $R(|\nabla \mathfrak{e}|^{2})\leq 2R({%
\mathfrak{e}})$. Note that $-\Delta _{\Gamma }=h(\mathfrak{e}_{\mathrm{Lapl}%
})$ and $\min \mathfrak{e}_{\mathrm{Lapl}}(\Gamma ^{\ast })=0$, where 
\begin{equation*}
\mathfrak{e}_{\mathrm{Lapl}}(p)\doteq 2\sum_{i=1}^{d}\left( 1-\cos
(p_{i})\right) 
\end{equation*}
is a dispersion fulfilling (\textbf{M}) with $R(\mathfrak{e}_{\mathrm{Lapl}%
})=1$.

To consider more general dispersions than $\mathfrak{e}_{\mathrm{Lapl}}$ is
important, for instance, for the analysis of many-body problems on the
lattice $\Gamma $ -- even in the situation where the dispersion relation for
the one-body sector is chosen to be $\mathfrak{e}_{\mathrm{Lapl}}$: Let $%
\mathfrak{e}$ be a dispersion relation. For each $K\in \Gamma ^{\ast }$
define the non-negative function $\mathfrak{e}^{(K)}:\Gamma
^{\ast}\rightarrow \mathbb{R}_{0}^{+}$ by 
\begin{equation}
\mathfrak{e}^{(K)}(p) \ = \ \mathfrak{e}\left( p\right) + \mathfrak{e}%
\left(K-p\right) - E_{0}^{(K)}\text{ },  \label{def eK}
\end{equation}%
where 
\begin{equation*}
E_{0}^{(K)} \ \doteq \ \min_{p^{\prime }\in \Gamma ^{\ast }} \left\{ 
\mathfrak{e}\left(p^{\prime }\right) + \mathfrak{e}\left( K-p^{\prime
}\right) \right\} \text{ }. 
\end{equation*}
Dispersions of the form (\ref{def eK}) come about in the analysis of systems
of two particles on the lattice $\Gamma $ both having the same dispersion $%
\mathfrak{e}$ and interacting by a (translation invariant) potential $%
V(x_{1}-x_{2})$. Indeed, the two-particle Hamiltonian is unitarily
equivalent to the direct integral 
\begin{equation*}
\int_{\Gamma ^{\ast }}^{\oplus }[H(\mathfrak{e}^{(K)},V)+E_{0}^{(K)}]\,%
\mathrm{d}\mu ^{\ast }(K)\text{ }. 
\end{equation*}%
The function $\mathfrak{e}^{(K)}$ is viewed as the (effective) dispersion of
a pair of particles travelling through the lattice with total quasi-momentum 
$K\in \Gamma ^{\ast }$. Clearly, $\mathfrak{e}^{(K)}$ fulfills (\textbf{M})%
\textbf{\ }-- at least in a neighborhood of $K=0$ --, if $\mathfrak{e}$
does. As soon as $K\neq 0$, however, $\mathfrak{e}_{\mathrm{Lapl}}^{(K)}$ is
not proportional to $\mathfrak{e}_{\mathrm{Lapl}}$. Similar facts hold true
for the $N$-body problem, $N>2$.

Our aim in the current paper is to give bounds on the size $N_{\mathrm{pp}}[%
\mathfrak{e},V]$ of the pure point spectrum of $H(\mathfrak{e},V)$, 
\begin{equation*}
N_{\mathrm{pp}}[\mathfrak{e},V]\ \doteq \mathrm{Tr}\left[ E_{\mathrm{pp}}(H(%
\mathfrak{e},V))\right] \text{ }, 
\end{equation*}
in dimensions $d\geq 3$. Here, $E_{\mathrm{pp}}(H)$ is the spectral
projector associated with the pure-point spectrum of $H(\mathfrak{e},V)$,
i.e., the range of $E_{\mathrm{pp}}(H)$ is the closed linear span of the
eigenvectors of the selfadjoint operator $H$. In quantum mechanics, $N_{%
\mathrm{pp}}[\mathfrak{e},V]$ is the number of linearly independent bound
states of a particle moving in the $d$-dimensional lattice $\Gamma $, with
dispersion $\mathfrak{e}$ and in presence of the potential $V$. Another
important physical aspect of the quantity $N_{\mathrm{pp}}[\mathfrak{e},V]$
concerns the scattering of such a particle on the potential $V$: When the
particle is not bound by the potential (i.e., its energy distribution
vanishes at all eigenvalues of the Hamiltonian $H(\mathfrak{e},V)$) then $N_{%
\mathrm{pp}}[\mathfrak{e},V]$ is related to the time-delay caused by the
scattering process. See, for instance, \cite[Eq. (1)]{BS-B12} and references
therein.

By a theorem of von Neumann and Weyl \cite[Chapter X, Theorem 2.1]{Kato},
for any self-adjoint operator $H_{0}$ on a separable Hilbert space $\mathcal{%
H}$ and any prescribed upper bound $\varepsilon >0$, there is another
self-adjoint operator $H_{1}$ with $\dim E_{\mathrm{pp}}(H_{1})=\infty $ and
$H_{1}-H_{0}$ smaller than $\varepsilon $ in the Hilbert-Schmidt norm. Thus,
even arbitrarily small perturbations can change the pure-point spectrum of a
self-adjoint operator drastically. The appearance of infinitely many
eigenvalues driven by an arbitrarily small perturbation is also known for
the special case of lattice Schr\"{o}dinger operators: If $d=1$ and $%
\mathfrak{e}=\mathfrak{e}_{\mathrm{Lapl}}$ then, for any $\varepsilon >0$,
there is a potential $V_{\varepsilon }$ such that $|V_{\varepsilon }(x)|\leq
\varepsilon (1+|x|)^{-1}$ and the eigenvalues of $H(\mathfrak{e}%
,V_{\varepsilon })$ are dense in the interval $[0,{\mathfrak{e}_{\mathrm{max}%
}}(\mathfrak{e})]=[0,4]$. See \cite[Theorem 2.1]{Na}. In particular, $N_{%
\mathrm{pp}}[\mathfrak{e},V_{\varepsilon }]=\infty$, whereas $N_{\mathrm{pp}%
}[\mathfrak{e},0]=0$. Note that $[0,{\mathfrak{e}_{\mathrm{max}}}(\mathfrak{e%
})]$ is exactly the essential spectrum of $H(\mathfrak{e},V_{\varepsilon })$
and thus arbitrarily small perturbations of the potential of a discrete Schr%
\"{o}dinger operator can even generate infinitely many \textit{embedded}
eigenvalues. The appearance of embedded eigenvalues is related to the slow
decay of potentials:\ As proven in \cite{imb}, if $d=1$, $\mathfrak{e}=%
\mathfrak{e}_{\mathrm{Lapl}}$ and, for some $\eta >0$ and $C<\infty $, $%
|V(x)|\leq C(1+|x|)^{-(1+\eta )}$ then $H(\mathfrak{e},V)$ has no eigenvalue
in the (open) interval $(0,{\mathfrak{e}_{\mathrm{max}}}(\mathfrak{e}))$.
However, to our knowledge, the precise relation between the presence of
embedded eigenvalues and the slow decay of the potential $V$ has not yet
been established for discrete Schr\"{o}dinger operators in dimension higher
than one. Moreover, this property of embedded eigenvalues of discrete Schr%
\"{o}dinger operators strongly depends (even in one dimension) on the choice 
$\mathfrak{e}$ of the dispersion, as one sees in the following simple
example:\ For $d \in \mathbb{N}$, define the dispersion 
\begin{equation*}
\mathfrak{e}(p) \ \doteq \ \frac{3d}{2} - \sum_{k=1}^{d}\left[ 2\cos(p_{k}))
- \cos (2p_{k}) \right] \ \geq \ 0 \text{ }. 
\end{equation*}
With this particular choice, ${\mathfrak{e}_{\mathrm{max}}}(\mathfrak{e}%
)=9d/2$. Note, moreover, that $\mathfrak{e}$ a dispersion satisfying (%
\textbf{M}). Further, let $\psi \in \ell^{2}(\Gamma )$ be defined by $%
\psi(x) \doteq (1 + |x|)^{-(d+1)/2}$. With this definition one has: 
\begin{equation*}
\left[ (h(\mathfrak{e})- 3d/2)\psi \right](x) \ = \ \mathcal{O}\Big( \big( %
1+|x| \big)^{-2-(d+1)/2} \Big) \text{ }. 
\end{equation*}
Define next the (real-valued) potential $V$ by 
\begin{equation*}
V(x) \ \doteq \ -\frac{\left[ (h(\mathfrak{e}) - 3d/2) \psi \right](x) }{
\psi(x)} \ = \ \mathcal{O}\Big( \big( 1+|x| \big)^{-2} \Big) \text{ }. 
\end{equation*}
This asymptotics is related to the fact that $\mathfrak{e}(p) - 3d/2 = 
\mathcal{O}\left( |p|^{2} \right)$. By construction, 
\begin{equation*}
\big[ \big( h(\mathfrak{e}) - 3d/2 \big)\psi \big] (x) \ = \ - V(x) \psi(x)%
\text{ }. 
\end{equation*}
In particular, $3d/2$ is an embedded eigenvalue of $H(\mathfrak{e},V)$.
Simple variations of this construction permit us to obtain potentials $V$ as
small as desired, producing embedded eigenvalues and decaying as $x^{-2}$.
Relaxing the condition (\textbf{M}) and admitting dispersions (which are not
Morse functions) such that $\mathfrak{e}(p)-c=\mathcal{O}\left(
|p|^{m}\right) $ for some $c>0$ and $m>2$, one can explicitly construct
arbitrarily small potentials $V(x)=\mathcal{O}\left( |x|^{-m}\right)$
leading to embedded eigenvalues for $H(\mathfrak{e},V)$.

Note that if the dispersion $\mathfrak{e}$ is a trigonometric polynomial
(i.e., $h(\mathfrak{e})$ has finite range) and $V$ has a finite support,
then $H(\mathfrak{e},V)$ has no embedded eigenvalues, i.e., no eigenvalue in
the interval $(0,{\mathfrak{e}}_{\mathrm{max}}(\mathfrak{e}))$ \cite[%
Proposition 9]{BS-B12}. In this particular case, the study of the size of
the pure point spectrum reduces to the study of the size of the discrete
spectrum (i.e., eigenvalues away from the essential spectrum $[0,{\mathfrak{e%
}}_{\mathrm{max}}(\mathfrak{e})]$) of $H(\mathfrak{e},V)$ and multiplicities
of possible (threshold) eigenvalues $0$ and ${\mathfrak{e}}_{\mathrm{max}}(%
\mathfrak{e})$. Observe, however, that even for zero-range potentials
threshold eigenvalues cannot be excluded, as shown in the following example:
Let $d\geq 5$ and define the potential 
\begin{equation*}
V(x)\doteq -\delta _{x,0}\left[ \int_{\Gamma ^{\ast }}\frac{1}{\mathfrak{e}_{%
\mathrm{Lapl}}(p)}\mathrm{d}\mu ^{\ast }(p)\right] ^{-1}, 
\end{equation*}
where $\delta _{x,y}$ is the Kronecker delta. Note that the above integral
is finite if $d\geq 3$. Further, define $\psi \in \ell^{2}(\Gamma )$ by 
\begin{equation*}
\psi(x) \ = \ \int_{\Gamma ^{\ast }} \frac{\mathrm{e}^{ip\cdot x}}{\mathfrak{%
e}_{\mathrm{Lapl}}(p)} \mathrm{d}\mu ^{\ast }(p)\text{ }, 
\end{equation*}
i.e., $\psi $ is the inverse Fourier transform of $\mathfrak{e}_{\mathrm{Lapl%
}}^{-1}$, the latter being an element of $L^{2}(\Gamma ^{\ast })$ for $d\geq
5$. Then, by construction, 
\begin{equation*}
\left[ h(\mathfrak{e}_{\mathrm{Lapl}})\psi \right](x) \ = \ -V(x)\psi (x)%
\text{ } . 
\end{equation*}
In particular $0$ is an (threshold) eigenvalue of $H(\mathfrak{e}_{\mathrm{%
Lapl}},V)$. See also \cite{HSSS12}.

From the discussion above one sees that the behavior of the dispersion $%
\mathfrak{e}$\ at critical points, as well as, the decay of the potential $V$
have a strong influence on the embedded eigenvalues of $H(\mathfrak{e},V)$.
As already explained above, we assume that the dispersion $\mathfrak{e}$
satisfies the condition (\textbf{M}). Later on, for technical reasons, we 
additionally assume that $\mathfrak{e} \in C^{4}(\Gamma^{\ast })$. In order
to control the spatial decay of the potential we define the following
quantity: For any $m \in \mathbb{N}$, $n\in \mathbb{N}_{0}$, and potential $%
V: \Gamma \rightarrow {\mathbb{R}}$, 
\begin{equation}
\Phi _{m,n}(V) \ \doteq \ \left( \sum\limits_{x\in \Gamma } |V(x)|^{\frac{1}{%
m}} \, (|x|+1)^{n} \right)^{m}\text{ }.  \label{def.Phi}
\end{equation}
If $\Phi _{1,2}(V) < \infty$ then $H(\mathfrak{e},V)$ has no singular
continuous spectrum and its pure point spectrum, is finite on any compact
subset of the real line not containing any critical value of the dispersion $%
\mathfrak{e}$. See Corollary \ref{koro.no.sing.spec}. In particular, the
eigenvalues of $H(\mathfrak{e},V)$ cannot be dense in its essential spectrum
in this case. Nevertheless, $N_{\mathrm{pp}}[\mathfrak{e},V]$ could still be
infinite, as its eigenvalues may possibly accumulate at critical values of $%
\mathfrak{e}$.

We prove below that, for a finite constant $c(\mathfrak{e})$ depending only
on a few derivatives of $\mathfrak{e}$, 
\begin{equation}
N_{\mathrm{pp}}[\mathfrak{e},V] \ \leq \ \inf\Big\{ |\mathrm{supp}\text{ }%
V^{\prime }| \text{ } \Big| \text{ } z \in \mathbb{Z}^{d} \text{ },\text{ }%
V^{\prime }\text{ with } \Phi _{2,3}(V^{(z)}-V^{\prime })<c(\mathfrak{e}) %
\Big\} \text{ },  \label{main bound Npp}
\end{equation}
where $|\mathrm{supp}$ $V^{\prime }|$ denotes the cardinality of the support 
$\mathrm{supp}$ $V^{\prime }\subset \Gamma $ of the potential $V^{\prime }$,
and $V^{(z)}$ is the translated potential 
\begin{equation}
V^{(z)}(x) \ \doteq \ V(z+x)\text{ }.  \label{transla V}
\end{equation}
This is our main result; see Theorem \ref{th main}. An immediate consequence
of this estimate is that $N_{\mathrm{pp}}[\mathfrak{e},V]$ is finite
whenever $\Phi _{2,3}(V)$ is finite. The bound (\ref{main bound Npp})
follows from resolvent estimates, given in Theorem~\ref{lap}, combined with
positivity arguments in the form of a Virial theorem, formulated as Lemma~%
\ref{Virial}, for $H(\mathfrak{e},V)$.

Note that (\ref{main bound Npp}) implies that $H(\mathfrak{e},V)$ has no
eigenvalue whenever $\Phi _{2,3}(V)<c(\mathfrak{e})$. This result on the
absence of pure-point spectrum is slightly strengthened in Corollary~\ref%
{th.pur.cont} (i), which states that $N_{\mathrm{pp}}[\mathfrak{e},V]=0$
when $\Phi _{2,2}(V)<c(\mathfrak{e})$. Moreover, the estimate (\ref{main
bound Npp}) can be used to prove the absence of eigenvalues of $H(\mathfrak{e%
},V)$ in its continuos spectrum, even if $\Phi _{2,3}(V)$ is big (but
finite): If $H(\mathfrak{e},V)$ has $N$ \textit{discrete} (isolated)
eigenvalues, counting their multiplicities, and $\Phi
_{2,3}(V^{(z)}-V^{\prime })<c(\mathfrak{e})$, for some $z\in \mathbb{Z}^{d}$
and some potential $V^{\prime }$ with $|\mathrm{supp}$ $V^{\prime }|=N$,
then it directly follows from (\ref{main bound Npp}) that $H(\mathfrak{e},V)$
has only discrete eigenvalues. See Corollary \ref{koro.embedded.ev}. 
Observe that, by Kato's perturbation theory for discrete eigenvalues \cite%
{Kato}, one sufficient condition on $V$ in order that $H(\mathfrak{e},V)$
has $N$ discrete eigenvalues is that there exists some potential $V^{\prime }
$, such that $|\mathrm{supp}$ $V^{\prime }|=N$ and%
\begin{equation*}
\left\Vert H(\mathfrak{e},V-V^{\prime })\right\Vert _{\mathcal{B}[\ell
^{2}(\Gamma )])}\left[ \min_{x\in \mathrm{supp}V^{\prime }}|V^{\prime }(x)|%
\right] ^{-1}
\end{equation*}%
is small enough.

Assume that $V$ is a potential with%
\begin{equation*}
|V(x)|\leq (1+|x|)^{-\beta } 
\end{equation*}%
for some $\beta >0$. If $\beta >2(d+3)$ then $\Phi _{2,3}(\lambda V)<\infty $
and, hence, $N_{\mathrm{pp}}[\mathfrak{e},\lambda V]$ is finite for all $%
\lambda \in \mathbb{R}$. In this case the estimate (\ref{main bound Npp})
yields 
\begin{equation*}
N_{\mathrm{pp}}[\mathfrak{e},\lambda V]=\mathcal{O}(\lambda ^{\frac{d}{\beta
-2(d+3)}})\text{ }. 
\end{equation*}%
On the other hand, if 
\begin{equation*}
V(x)\leq -(1+|x|)^{-\beta }\text{ }, 
\end{equation*}%
by estimating the size of the negative part of the pure-point spectrum of $H(%
\mathfrak{e},V)$ (i.e., its discrete spectrum) via the min-max principle,
one concludes that $N_{\mathrm{pp}}[\mathfrak{e},\lambda V]$ cannot be $%
o(\lambda ^{\frac{d}{\beta }})$. See also \cite{BdSPL10, RS08, RS09}. Ergo,
the estimate (\ref{main bound Npp}) is order-sharp for large, fast decaying
potentials.

Finally, observe that in \cite{BdSPL10} we proved a variational principle
similar to (\ref{main bound Npp}) for the size $N_{\mathrm{disc}}[\mathfrak{e%
},V]\leq N_{\mathrm{pp}}[\mathfrak{e},V]$ of the discrete spectrum of $H(%
\mathfrak{e},V)$, in any dimension $d\geq 1$, but for potentials $V$ having
a definite sign: In the case of the discrete spectrum, by using the
Birman-Schwinger principle, bounds like (\ref{main bound Npp}) on $N_{%
\mathrm{d}}[\mathfrak{e},V]$ can be obtained whereby the quantity $\Phi
_{2,3}(V)$ is replaced with $\Phi _{1,n}(V)$, $n$ being a small integer
depending on the dimension. For the case of the pure-point spectrum
considered in the current paper, however, that method is not applicable, for
the Birmann-Schwinger principle does not capture embedded eigenvalues (at
least not directly).

This paper is organized as follows:

\begin{itemize}
\item In Section \ref{spectrum.gen} we discuss general facts about the
spectrum of $H(\mathfrak{e},V)$ and prove a Virial theorem for $H(\mathfrak{e%
},V)$, as Lemma \ref{Virial}, which is pivotal for the proof of the estimate
(\ref{main bound Npp}).

\item In Section \ref{resolv.estmates} we derive resolvent estimates leading
to a limiting absorption principle (Theorem \ref{lap}), which is a central
ingredient of the proof of (\ref{main bound Npp}). An important technical
problem we are facing in these estimates arises from singularities of the
type%
\begin{equation*}
\frac{1}{p_{1}^{2}+\cdots +p_{k}^{2}-p_{k+1}^{2}\cdots -p_{d}^{2}} 
\end{equation*}%
appearing in integrands. Such singularities are called \textit{van Hove
singularities} in condensed matter physics and have important physical
consequences. They cannot be handled by simple \textit{power counting}, and
rather sign cancellations have to be exploited in the bounds. This technical
aspect is discussed in detail in Section \ref{proof.resolv} of the Appendix.

\item In Section \ref{sect.size.pp} we state and prove our main result,
Theorem \ref{th main}. Moreover, a result (Corollary \ref{koro.embedded.ev})
on the absence of embedded eigenvalues is derived from this last theorem.

\item To simplify the exposition and/or for completeness, some technical
results are proven in the Appendix (Section \ref{Appendix}).
\end{itemize}

\section{The Spectrum of $H(\mathfrak{e},V)$ -- General Facts\label%
{spectrum.gen}}

\noindent We require that $V$ decays at infinity, 
\begin{equation*}
V\ \in \ \ell _{0}^{\infty }(\Gamma ;\mathbb{R})\ \doteq \ \left\{ V:\Gamma
\rightarrow \mathbb{R}\;\Big|\ \lim_{|x|\rightarrow \infty }V(x)=0\right\} 
\text{ }, 
\end{equation*}%
or sometimes even that $V$ has finite support. Note that $V\in \ell
_{0}^{\infty }(\Gamma ;\mathbb{R})$ is compact as a multiplication operator
on $\ell ^{2}(\Gamma )$ and by a theorem of Weyl, 
\begin{equation*}
\sigma _{\mathrm{ess}}[H(\mathfrak{e},V)]\ =\ \sigma _{\mathrm{ess}}[H(%
\mathfrak{e},0)]\ =\ [0,{\mathfrak{e}_{\mathrm{max}}}]\text{ }, 
\end{equation*}%
where {$\mathfrak{e}$}${_{\mathrm{max}}}\equiv \,${$\mathfrak{e}$}${_{%
\mathrm{max}}}(\mathfrak{e})$ and $\sigma _{\mathrm{ess}}[H]\subset {\mathbb{%
R}}$ denotes the essential spectrum of the selfadjoint operator $H$.

Let {$\mathfrak{e}\in $}$C^{2}(\Gamma ^{\ast };{\mathbb{R}})$ be a Morse
function. As $\Gamma ^{\ast }$ is compact, {$\mathfrak{e}$ has} at most
finitely many critical points. We denote the set of all critical points of {$%
\mathfrak{e}$} by 
\begin{equation*}
\mathrm{Crit}({\mathfrak{e}})\doteq \{p\in \Gamma ^{\ast }\;|\;\nabla {%
\mathfrak{e}}(p)=0\}\text{ }. 
\end{equation*}%
The critical values of {$\mathfrak{e}$, collected in the }set 
\begin{equation*}
\mathrm{Thr}({\mathfrak{e}})\doteq {\mathfrak{e}}\left( \mathrm{Crit}({%
\mathfrak{e}})\right) \text{ }, 
\end{equation*}%
are called of \textit{thresholds} of ${\mathfrak{e}}$.

Define the symmetric operator $\hat{A}=\hat{A}({\mathfrak{e}})$ on $%
C^{\infty }(\Gamma ^{\ast };{\mathbb{C}})\subset L^{2}(\Gamma ^{\ast })$ by 
\begin{equation}
\hat{A}{\hat{\psi}}(p)\doteq i\sum\limits_{i=1}^{d}\left\{ [\partial _{p_{i}}%
{\mathfrak{e}}(p)][\partial _{p_{i}}{\hat{\psi}}(p)]+\frac{1}{2}[\partial
_{p_{i}}^{2}{\mathfrak{e}}(p)]{\hat{\psi}}(p)\right\} \text{ }.
\label{Def.A}
\end{equation}%
We denote by $A=A({\mathfrak{e}})$ the inverse Fourier transform of $\hat{A}$%
, i.e., the operator $A=\mathcal{F}\,\hat{A}\,\mathcal{F}^{\ast }$ on $%
\mathrm{Dom}(A)=\mathcal{F}(C^{\infty }(\Gamma ^{\ast };{\mathbb{C}}))$.
Observe that, for all ${\hat{\psi}}\in C^{\infty }(\Gamma ^{\ast };{\mathbb{C%
}})$, 
\begin{equation}
i[\mathfrak{e},\hat{A}]{\hat{\psi}}(p)=|\nabla {\mathfrak{e}}(p)|^{2}\hat{%
\psi}(p)\doteq \sum\limits_{i=1}^{d}[\partial _{p_{i}}{\mathfrak{e}}(p)]^{2}{%
\hat{\psi}}(p)\text{ }.  \label{pos.commut}
\end{equation}
In particular, $i[\mathfrak{e},\hat{A}]$ and $i[h({\mathfrak{e}}),A]$ extend
to positive bounded operators on $L^{2}(\Gamma ^{\ast })$ and $\ell
^{2}(\Gamma )$, which we also denote a by $i[{\mathfrak{e}},\hat{A}]$ and $%
i[h({\mathfrak{e}}),A]$, respectively. Note also that $|\nabla {\mathfrak{e}}%
(p)|^{2}$ is a Morse function, by Assumption (\textbf{M}).

Note that $i[V,A]$ uniquely extends to a bounded self-adjoint operator on $%
\ell ^{2}(\Gamma )$ (also denoted by $i[V,A]$) whenever $V$ has a finite
support. (Below, densely defined bounded operators will be identified with
their closures.) More precisely, from straightforward computations one
obtains 
\begin{equation}
i[V,A]\ =\ \sum\limits_{x\in \mathrm{supp}\text{ }V}iV(x):\Big(|\delta
_{x}\rangle \langle g_{x}|-|g_{x}\rangle \langle \delta _{x}|\Big)\text{ },
\label{va}
\end{equation}%
where $g_{x}\doteq A\delta _{x}$ has Fourier transform 
\begin{equation*}
\hat{g}_{x}(p)\doteq \big[\frac{i}{2}|\nabla _{p}\mathfrak{e}%
(p)|^{2}+\langle \nabla _{p}\mathfrak{e}(p),x\rangle \big]\,\mathrm{e}%
^{-i\langle p,x\rangle }
\end{equation*}
and obeys thus the norm bound 
\begin{equation*}
\Vert g_{x}\Vert _{\ell ^{2}(\Gamma )}\ =\ \Vert \hat{g}_{x}\Vert
_{L^{2}(\Gamma ^{\ast })}\ \leq \ C\big(1+|x|\big)\text{ }.
\end{equation*}%
In particular, it follows from (\ref{va}) that, if $V(x)\,(1+|x|)$ is
summable, then $\lim_{R\rightarrow \infty }\Vert \mathbf{1}_{|x|>R}\,[V,A]\,%
\mathbf{1}_{|x|>R}\big\|=0$, and hence $i[V,A]$ is a compact operator. For
such potentials we have the following estimate for the commutator $i[H({%
\mathfrak{e}},V),A]$:

\begin{lemma}[Mourre Estimate for $H({\mathfrak{e}},V)$]
\label{Mourre} If ${\mathfrak{e}}\in C^{4}(\Gamma ^{\ast };{\mathbb{R}})$ is
a dispersion relation then $A({\mathfrak{e}})$ uniquely extends to a
self-adjoint operator (also denoted by $A({\mathfrak{e}})$). If the
potential $V$ is such that $i[V,A]$ is compact then, for any continuous,
compactly supported function $\chi :{\mathbb{R}}\rightarrow {\mathbb{R}}$
satisfying $\mathrm{dist}(\mathrm{Thr}\{ {\mathfrak{e}}),\, \mathrm{supp}\;
\chi \} > 0$, there is a compact operator $K_{\chi }\in \mathcal{B}[\ell
^{2}(\Gamma )]$ and a constant $c_{\chi }>0$ such that 
\begin{equation}
\chi \left[ H({\mathfrak{e}},V)\right] \,i[H({\mathfrak{e}},V),A]\,\chi %
\left[ H({\mathfrak{e}},V)\right] \geq c_{\chi }\chi ^{2}\left[ H({\mathfrak{%
e}},V)\right] +K_{\chi }\text{ }.  \label{Moure estimate}
\end{equation}
\end{lemma}

Observe that if $\Delta \subset {\mathbb{R}}$ is a compact subset with $%
\mathrm{dist}\{\mathrm{Thr}({\mathfrak{e}}),\Delta \}>0$, then there is a
continuous function $\chi :{\mathbb{R}}\rightarrow {\mathbb{R}}$ with
compact support such that $\mathrm{dist}\{\mathrm{Thr}({\mathfrak{e}}), 
\mathrm{supp}\; \chi\}>0$, and $\chi \equiv 1$ on $\Delta$. Let $E_{\Delta }$
be the spectral projection of $H({\mathfrak{e}},V)$ associated with $\Delta $%
. Then $\chi E_{\Delta }=\chi E_{\Delta }=E_{\Delta }$ and by multiplying
equation (\ref{Moure estimate}) with $E_{\Delta }$ from the left and from
the right it follows that, for some $c_{\Delta }>0$ and some compact
operator $K_{\Delta }$, 
\begin{equation}  \label{Mourre-estimate-delta}
E_{\Delta }i[H({\mathfrak{e}},V),A]E_{\Delta }\geq c_{\Delta }E_{\Delta
}+K_{\Delta }\text{ }.
\end{equation}
From explicit expressions for $AAV$, $AVA$ and $VAA$, similar to (\ref{va}), one checks that these three operators are bounded if $\Phi
_{1,2}(V)<\infty $. By (\ref{pos.commut}), if ${\mathfrak{e}}\in
C^{3}(\Gamma ^{\ast };{\mathbb{R}})$, $[[h({\mathfrak{e}}),A],A]$ is a
bounded operator. In particular, if $\Phi _{2,2}(V)<\infty $ and ${\mathfrak{%
e}}\in C^{3}(\Gamma ^{\ast };{\mathbb{R}})$, then 
\begin{equation}  \label{Mourre-estimate-double-commutator}
\big\| \: [[H({\mathfrak{e}},V),A],A] \: \big\|_{\mathcal{B}[\ell^{2}(\Gamma
)]} \ < \ \infty \text{ }.
\end{equation}

The following corollary is a consequence of (\ref{Mourre-estimate-delta})
and (\ref{Mourre-estimate-double-commutator}); see also \cite[Theorems 4.7
and 4.9]{Cycon}.

\begin{corollary}
\label{koro.no.sing.spec} Let ${\mathfrak{e}}\in C^{4}(\Gamma ^{\ast };{%
\mathbb{R}})$ be a dispersion relation and let $V$ be a potential with $\Phi
_{1,2}(V)<\infty $. Then $H({\mathfrak{e}},V)$ has no singular continuous
spectrum and its eigenvalues can only accumulate in points of $\mathrm{Thr}({%
\mathfrak{e}})$.
\end{corollary}

The next lemma, along with Theorem \ref{lap}, is a central argument of the
proof of our main result (Theorem \ref{th main}):

\begin{lemma}[Virial Theorem for $H({\mathfrak{e}},V)$]
\label{Virial} Let ${\mathfrak{e}}\in C^{4}(\Gamma ^{\ast };{\mathbb{R}})$
be a dispersion relation and let $V_{1},V_{2}$ be potentials such that $%
i[V_{1},A({\mathfrak{e}})]$ and $i[V_{2},A({\mathfrak{e}})]$ are bounded
operators. If $\psi $ is an eigenvector of $H({\mathfrak{e}},V_{1}+V_{2})$,
then 
\begin{equation*}
\langle \psi \,|\,i[V_{2},A]\psi \rangle \ = \ -\langle \psi \,|\,i[H({%
\mathfrak{e}},V_{1}),A]\psi \rangle \text{ }. 
\end{equation*}
\end{lemma}

\noindent Note that, in the current section, the restriction ${\mathfrak{e}}
\in C^{4}(\Gamma ^{\ast };{\mathbb{R}})$ is only relevant for Corollary~\ref%
{koro.no.sing.spec} and Lemma~\ref{Virial} above. The proofs of Lemmata~\ref%
{Mourre} and \ref{Virial} use adaptations for the lattice case of known
methods used for the continuum and are given in Appendix~\ref{proof.moure}--%
\ref{proof.virial}, for completeness. See also \cite[Chapter 4]{Cycon} and 
\cite{Graf}.

The following upper bound on the on the size of the pure-point spectrum of $%
H({\mathfrak{e}},V)$, in case that $V$ has finite support and $h({\mathfrak{e%
}})$ is of finite range, is an immediate consequence of the Virial theorem
(Lemma~\ref{Virial}) above:

\begin{corollary}[{Upper Bound on $N_{pp}[\mathfrak{e},V]$, Finite Range Case%
}]
\label{koro.fin.ren.u.b} Let $d\geq 1$. If the potential $V$ has finite
support and ${\mathfrak{e}}\in C^{4}(\Gamma ^{\ast };{\mathbb{R}})$ is a
dispersion relation satisfying (\textbf{M}) then 
\begin{equation*}
N_{\mathrm{pp}}[\mathfrak{e},V]\leq \mathrm{Tr}\left[ E_{-}\left( i[V,A({%
\mathfrak{e}})]\right) \right] \text{ },
\end{equation*}%
where $E_{-}(i[V,A])$ is the spectral projector associated with the \emph{%
strictly} negative spectrum of the (selfadjoint) commutator $i[V,A]$.
\end{corollary}

\noindent \textit{Proof.} Note that the range of $i[V,A]$ has finite
dimension, by (\ref{va}). From (\ref{pos.commut}), $i[h({%
\mathfrak{e}}),A]$ is positive and has purely absolutely continuous
spectrum. Thus, for all $\psi \in \ell ^{2}(\Gamma )\backslash \{0\}$, $%
\langle \psi \,|\,i[h({\mathfrak{e}}),A]\psi \rangle >0$. Setting $V_{1}=0$
and $V_{2}=V$, it follows from Lemma \ref{Virial} that, for any normalized
eigenvector $\psi $ of $H({\mathfrak{e}},V)$, we have%
\begin{equation*}
\langle \psi \,|\,i[V,A]\psi \rangle =-\langle \mathcal{F}^{\ast }(\psi )\,|%
\text{ }|\,\nabla \mathfrak{e}|^{2}\mathcal{F}^{\ast }(\psi )\rangle <0\text{
}. 
\end{equation*}

Hence, denoting by $X\subset \ell ^{2}(\Gamma )$ any finite-dimensional
subspace of eigenvectors of $H({\mathfrak{e}},V)$ we obtain, by compactness
of the $n$-sphere, with $n=\mathrm{dim}X-1$, the estimate 
\begin{equation*}
\max \big\{\langle \psi |\,i[V,A]\psi \rangle \text{ }\big|\text{ }\psi \in
X,\text{ }\Vert \psi \Vert _{2}=1\big\}\ <\ 0\text{ }.
\end{equation*}%
By the min-max principle, the dimension of $X$ cannot exceed the number of 
\emph{strictly }negative eigenvalues (with multiplicities) of the
self-adjoint operator $i[V,A]$.\hfill $\square $\smallskip 

In the following corollary we show that the quantity $\mathrm{Tr}\left[
E_{-}\left( i[V,A({\mathfrak{e}})]\right) \right] $ (appearing in the above
estimate on $N_{\mathrm{pp}}[\mathfrak{e},V]$) is nothing else than the size
of the support of the potential $V$:

\begin{corollary}
\label{koro.fin.ren.u.b copy(1)} Let $d\geq 1$. If the potential $V$ has
finite support and ${\mathfrak{e}}\in C^{4}(\Gamma ^{\ast };{\mathbb{R}})$
is a dispersion relation satisfying (\textbf{M}) then 
\begin{equation*}
\mathrm{Tr}\left[ E_{-}\left( i[V,A({\mathfrak{e}})]\right) \right] =|%
\mathrm{supp}\text{ }V|\text{ }. 
\end{equation*}
\end{corollary}

\noindent \textit{Proof.} Note that, for all $\lambda >0$,%
\begin{equation*}
\mathrm{Tr}\left[ E_{-}\left( i[\lambda V,A({\mathfrak{e}})]\right) \right] =%
\mathrm{Tr}\left[ E_{-}\left( i[V,A({\mathfrak{e}})]\right) \right] \text{.}
\end{equation*}%
By Kato's perturbation theory for discrete eigenvalues \cite{Kato}, for
sufficiently large $\lambda >0$, $H({\mathfrak{e}},\lambda V)$ has, at
least, $|\mathrm{supp}$ $V|$ eigenvalues, counting their multiplicities.
Hence, Corollary~\ref{koro.fin.ren.u.b} implies that 
\begin{equation*}
|\mathrm{supp}\text{ }V|\leq N_{\mathrm{pp}}[\mathfrak{e},\lambda V]\leq 
\mathrm{Tr}\left[ E_{-}\left( i[\lambda V,A({\mathfrak{e}})]\right) \right] =%
\mathrm{Tr}\left[ E_{-}\left( i[V,A({\mathfrak{e}})]\right) \right] \text{ }.
\end{equation*}%
Repeating this argument for $-V$ we conclude that:%
\begin{equation*}
|\mathrm{supp}\text{ }V|\leq \mathrm{Tr}\left[ E_{-}\left( \pm i[V,A({%
\mathfrak{e}})]\right) \right] \text{ }.
\end{equation*}%
In other words, the subspaces associated to the \emph{strictly} negative and 
\emph{strictly} positive eigenvalues of the selfadjoint, finite-range
operator $i[V,A({\mathfrak{e}})]$ have both dimension of at least $|\mathrm{%
supp}$ $V|$. On the other hand, from (\ref{va}) we conclude that the
dimension of the range of this operator, which is%
\begin{equation*}
\mathrm{Tr}\left[ E_{-}\left( i[V,A({\mathfrak{e}})]\right) \right] +\mathrm{%
Tr}\left[ E_{-}\left( -i[V,A({\mathfrak{e}})]\right) \right] \text{ },
\end{equation*}%
cannot exceed $2|\mathrm{supp}$ $V|$. Ergo, 
\begin{equation*}
\mathrm{Tr}\left[ E_{-}\left( i[V,A({\mathfrak{e}})]\right) \right] =\mathrm{%
Tr}\left[ E_{-}\left( -i[V,A({\mathfrak{e}})]\right) \right] =|\mathrm{supp}%
\text{ }V|\text{ }.
\end{equation*}%
 $\ $\hfill $\square $\smallskip 

Combining the two last corollaries, under the same assumptions on the
dispersion ${\mathfrak{e}}$, we arrive at:%
\begin{equation*}
N_{\mathrm{pp}}[\mathfrak{e},V]\leq |\mathrm{supp}\text{ }V|\text{ }. 
\end{equation*}

In dimension $d\geq 3$, the upper bound on $N_{\mathrm{pp}}[\mathfrak{e},V]$
of Corollary \ref{koro.fin.ren.u.b} can be improved in the following sense:

\begin{itemize}
\item If $\Phi _{2,2}(V)$ is small enough then $H({\mathfrak{e}},V)$ has no
bound states cf. Corollary \ref{th.pur.cont} (i).

\item If $V=V_{1}+V_{2}$ with $V_{1}$ having finite support and $\Phi
_{2,3}(V_{2})$ being small enough (but $V$ not necessarily having a finite
support), then the bound on $N_{\mathrm{pp}}[\mathfrak{e},V]$ in the
corollary remains true when $V$ is replaced with $V_{1}$. See Corollary \ref%
{koro.u.b.N.2}.
\end{itemize}

\section{Resolvent Estimates\label{resolv.estmates}}

Let $\mathfrak{e}^{\prime \prime }(p)$ be the Hessian matrix of the
dispersion relation ${\mathfrak{e}}\in C^{2}(\Gamma ^{\ast };{\mathbb{R}})$
at $p\in \mathrm{Crit}({\mathfrak{e}})$. Define the \textit{minimal
curvature of $\mathfrak{e}$ at }$p\in \mathrm{Crit}({\mathfrak{e}})$ by: 
\begin{equation*}
K({\mathfrak{e}},p)\doteq \min \left\{ |\lambda |^{\frac{1}{2}}\;:\;\lambda
\,\text{is an eigenvalue of }\,\mathfrak{e}^{\prime \prime }(p)\right\} 
\text{ }. 
\end{equation*}%
Define also the \textit{minimal (critical) curvature of $\mathfrak{e}$} by 
\begin{equation*}
K({\mathfrak{e}})\doteq \min \{K({\mathfrak{e}},p)\;|\;p\in \mathrm{Crit}({%
\mathfrak{e}})\}\text{ }. 
\end{equation*}%
Note that $K({\mathfrak{e}})>0$ and $K(|\nabla {\mathfrak{e}}|^{2})>0$ under
Assumption (\textbf{M})\textbf{. }

For $m\in {\mathbb{N}}_{0}$, we recall the standard definition 
\begin{equation*}
\left\Vert {\mathfrak{e}} \right\Vert _{C^{m}} \ \doteq \ \max\limits_{ 
\underline{n} \in {\mathbb{N}}_{0}^{d} \newline
|\underline{n}|=m } \; \max\limits_{p \in \Gamma ^{\ast }} |\partial _{p}^{%
\underline{n}}{\mathfrak{e}}(p)| 
\end{equation*}
of the norm on $C^{m}(\Gamma^{\ast };{\mathbb{C}})$.

\begin{lemma}
\label{resolv} Let ${\mathfrak{e}}$ be any dispersion relation from $%
C^{3}(\Gamma ^{\ast };{\mathbb{R}})$. Let $K>0$ and $C<\infty $ be constants
with $K({\mathfrak{e}})\geq K$, and $\left\Vert {\mathfrak{e}}\right\Vert
_{C^{3}}\leq C$. Then there is a constant $c_{\ref{resolv}}<\infty $
depending only on $K$ and $C$ such that%
\begin{equation*}
\left\Vert V^{\frac{1}{2}}(z-h({\mathfrak{e}}))^{-1}V^{\frac{1}{2}%
}\right\Vert _{\mathcal{B}[\ell ^{2}(\Gamma )]} \ \leq \ c_{\ref{resolv}%
}\,\Phi _{2}(V)\text{ }, 
\end{equation*}

\begin{equation*}
\left. 
\begin{array}{l}
\left\Vert V^{\frac{1}{2}}(z-h({\mathfrak{e}}))^{-1}AV^{\frac{1}{2}%
}\right\Vert _{\mathcal{B}[\ell ^{2}(\Gamma )]} \\ 
\\ 
\left\Vert V^{\frac{1}{2}}A(z-h({\mathfrak{e}}))^{-1}V^{\frac{1}{2}%
}\right\Vert _{\mathcal{B}[\ell ^{2}(\Gamma )]} \\ 
\\ 
\left\Vert V^{\frac{1}{2}}A(z-h({\mathfrak{e}}))^{-1}AV^{\frac{1}{2}%
}\right\Vert _{\mathcal{B}[\ell ^{2}(\Gamma )]}%
\end{array}%
\right\} \ \leq \ c_{\ref{resolv}}\,\Phi _{3}(V)\text{ }, 
\end{equation*}

\begin{equation*}
|\langle \delta_{x}|(z-h({\mathfrak{e}}))^{-1} \delta_{y} \rangle | \ \leq \
c_{\ref{resolv}}^{2}(1+|x|)^{2}(1+|y|)^{2}\text{ }, 
\end{equation*}

\begin{equation*}
\left\Vert V^{\frac{1}{2}}(z-h({\mathfrak{e}}))^{-1}\varphi _{x}
\right\Vert_{2} \ \leq \ c_{\ref{resolv}}\,(1+|x|)^{2}\Phi _{2}(V)^{\frac{1}{%
2}}\text{ }, 
\end{equation*}

\begin{equation*}
\left\Vert V^{\frac{1}{2}}A(z-h({\mathfrak{e}}))^{-1}\varphi
_{x}\right\Vert_{2} \ \leq \ c_{\ref{resolv}}\,(1+|x|)^{2}\Phi _{3}(V)^{%
\frac{1}{2}}\text{ }, 
\end{equation*}%
for all potentials $V$, all $z\in \mathbb{C}\backslash \mathbb{R}$, and all $%
x,y\in \Gamma $. Here, $V^{\frac{1}{2}}$ denotes an arbitrary function $%
V:\Gamma \rightarrow \mathbb{C}$ with $(V^{\frac{1}{2}}(x))^{2}=V(x)$.
\end{lemma}

\noindent \textit{Proof. }We freely use the equality $((V^{\frac{1}{2}%
})^{\ast })^{2}=(V^{\frac{1}{2}})^{2}=V$ in the sequel without further
mentioning. We write 
\begin{equation*}
\Big\langle \delta_{x} \, \Big| \; \big( z - h({\mathfrak{e}}) \big)^{-1}
\varphi _{y} \Big\rangle
\ = \ (1+|x|)^{2} \, (1+|y|)^{2} \, \int_{\Gamma ^{\ast }} \frac{F_{xy}(p)}{%
z-e(p)} \, \mathrm{d}\mu ^{\ast }(p) \text{ }, 
\end{equation*}%
where 
\begin{equation*}
F_{xy}(p) \ \doteq \ \frac{\mathrm{e}^{ip\cdot (x-y)}}{(1+|x|)^{2}(1+|y|)^{2}%
}\text{ }, 
\end{equation*}%
and note that $\sup\big\{ \left\Vert F_{xy} \right\Vert_{C^{2}} \ \big| \
x,y \in \Gamma \big\} < \infty$. Hence, it follows from Lemma \ref%
{estimates.Morse.func} that there is a constant $\mathrm{const}<\infty $,
such that 
\begin{equation*}
\Big| \Big\langle \delta_{x} \, \Big| \; \big( z - h({\mathfrak{e}}) \big)%
^{-1} \, \delta_{y} \Big\rangle \Big|
\ \leq \ \mathrm{const}(1+|x|)^{2}(1+|y|)^{2} \text{ } , 
\end{equation*}
for all $z\in \mathbb{C}\backslash \mathbb{R}$ and all $x,y\in \Gamma $.

Let $V$ be a potential with $\mathrm{Ran}(V^{\frac{1}{2}})\subset \mathrm{dom%
}(A)$. For all $\psi \in \ell ^{2}(\Gamma )$, we define the following
functions on $\Gamma ^{\ast }$, 
\begin{eqnarray*}
F_{V}^{\psi }(p) &\doteq &\mathcal{F}^{\ast }\circ V^{\frac{1}{2}}(\psi
)(p)=\sum\limits_{x\in \Gamma }\mathrm{e}^{-ip\cdot x}V^{\frac{1}{2}}(x)\psi
(x)\text{ }, \\
F_{AV}^{\psi }(p) &\doteq &\sum\limits_{i=1}^{d}i[\partial _{p_{i}}{%
\mathfrak{e}}(p)][\partial _{p_{i}}F_{V}^{\psi }](p)+\frac{i}{2}|\nabla {%
\mathfrak{e}}(p)|^{2}F_{V}^{\psi }(p)\text{ }.
\end{eqnarray*}
Then, for all $x\in \Gamma $, 
\begin{eqnarray*}
\big\langle (\bar{z} - h({\mathfrak{e}}) )^{-1} \, V^{\frac{1}{2}} \psi \, %
\big| \ \delta_{x} \big\rangle & = & \int_{\Gamma ^{\ast }} \frac{\overline{%
F_{V}^{\psi }(p)}\mathrm{e}^{-ip\cdot x}}{z-{\mathfrak{e}}(p)} \, \mathrm{d}%
\mu ^{\ast }(p)\text{ }, \\
\big\langle (\bar{z} - h({\mathfrak{e}}) )^{-1} \, A V^{\frac{1}{2}} \psi \, %
\big| \ \delta_{x} \big\rangle & = & \int_{\Gamma ^{\ast }} \frac{\overline{%
F_{AV}^{\psi }(p)} \, \mathrm{e}^{-ip\cdot x}}{z-{\mathfrak{e}}(p)} \, 
\mathrm{d}\mu ^{\ast }(p)\text{ } .
\end{eqnarray*}%
We note 
\begin{equation*}
\overline{F_{\#}^{\psi }(p)}\mathrm{e}^{-ip\cdot x}=(1+|x|)^{2}\left[
(1+|x|)^{-2}\overline{F_{\#}^{\psi }(p)}\mathrm{e}^{-ip\cdot x}\right] \text{
}, 
\end{equation*}%
where $\#$ denotes $V$ or $AV$, and observe that the $C^{2}$-norms of the
functions 
\begin{equation*}
p\mapsto (1+|x|)^{-2}\overline{F_{V}^{\psi }(p)}\mathrm{e}^{-ip\cdot
x},\quad p\mapsto (1+|x|)^{-2}\overline{F_{AV}^{\psi }(p)}\mathrm{e}%
^{-ip\cdot x} 
\end{equation*}%
are bounded by $\mathrm{const}\,\Phi _{2}(V)^{\frac{1}{2}}$ and $\mathrm{%
const\,}\Phi _{3}(V)^{\frac{1}{2}}$, $\mathrm{const}<\infty $, respectively,
uniformly in $x\in \Gamma $ and $\psi \in \ell ^{2}(\Gamma )$, $\left\Vert
\psi \right\Vert _{2}\leq 1$. It follows from Lemma \ref%
{estimates.Morse.func} that, for some constant $\mathrm{const}<\infty $, all 
$x\in \Gamma $, all $z\in \mathbb{C}\backslash \mathbb{R}$, and all $\psi
\in \ell ^{2}(\Gamma )$, with $\left\Vert \psi \right\Vert _{2}\leq 1$, 
\begin{eqnarray*}
\Vert V^{\frac{1}{2}}(z-h({\mathfrak{e}}))^{-1}\delta _{x}\Vert _{2} &\leq &%
\mathrm{const}\,(1+|x|)^{2}\Phi _{2}(V)^{\frac{1}{2}}\text{ }, \\
\Vert V^{\frac{1}{2}}A(z-h({\mathfrak{e}}))^{-1}\delta _{x}\Vert _{2} &\leq &%
\mathrm{const\,}(1+|x|)\,^{2}\Phi _{3}(V)^{\frac{1}{2}}\text{ }, \\
\sum_{x\in \Gamma }|\langle (z-h({\mathfrak{e}}))^{-1}V^{\frac{1}{2}}\psi
\,|\,V^{\frac{1}{2}}\delta _{x}\rangle |^{2} &\leq &\mathrm{const^{2}\,}\Phi
_{2}(V)^{2}\text{ }, \\
\sum_{x\in \Gamma }|\langle (z-h({\mathfrak{e}}))^{-1}AV^{\frac{1}{2}}\psi
\,|\,V^{\frac{1}{2}}\delta _{x}\rangle |^{2} &\leq &\mathrm{const^{2}}\,\Phi
_{3}(V)\Phi _{2}(V) \\
&\leq &\mathrm{const^{2}}\,\Phi _{3}(V)^{2}\text{ }.
\end{eqnarray*}%
Thus, 
\begin{eqnarray*}
\Vert V^{\frac{1}{2}}(z-h({\mathfrak{e}}))^{-1}V^{\frac{1}{2}}\Vert _{%
\mathcal{B}[\ell ^{2}(\Gamma )]} &\leq &\mathrm{const}\,\Phi _{2}(V)\text{ },
\\
\Vert V^{\frac{1}{2}}(z-h({\mathfrak{e}}))^{-1}AV^{\frac{1}{2}}\Vert _{%
\mathcal{B}[\ell ^{2}(\Gamma )]} &\leq &\mathrm{const}\,\Phi _{3}(V)\text{ },
\end{eqnarray*}%
for some $\mathrm{const}<\infty $, all $z\in \mathbb{C}\backslash \mathbb{R}$
and all $x\in \Gamma $. By taking adjoints, we further obtain 
\begin{equation}
\Vert V^{\frac{1}{2}}A(z-h({\mathfrak{e}}))^{-1}V^{\frac{1}{2}}\Vert _{%
\mathcal{B}[\ell ^{2}(\Gamma )]}\leq \mathrm{const}\,\Phi _{3}(V)\text{ }.
\end{equation}%
Similarly, it follows, for a suitable constant $\mathrm{const}<\infty $, all 
$z\in \mathbb{C}\backslash \mathbb{R}$, all $x\in \Gamma $, and all $\psi
\in \ell ^{2}(\Gamma )$, $\left\Vert \psi \right\Vert _{2}\leq 1$, that 
\begin{equation*}
\sum_{x\in \Gamma }|\langle (z-h({\mathfrak{e}}))^{-1}AV^{\frac{1}{2}}\psi
\,|\,AV^{\frac{1}{2}}\delta _{x}\rangle |^{2}\leq \mathrm{const^{2}}\,\Phi
_{2,3}(V)^{2}\text{ }. 
\end{equation*}%
Thus, 
\begin{equation*}
\Vert V^{\frac{1}{2}}A(z-h({\mathfrak{e}}))^{-1}AV^{\frac{1}{2}}\Vert _{%
\mathcal{B}[\ell ^{2}(\Gamma )]}\leq \mathrm{const}\,\Phi _{2,3}(V) 
\end{equation*}%
for all $z\in \mathbb{C}\backslash \mathbb{R}$ and all $x\in \Gamma $.
\hfill $\square $

\begin{theorem}[Resolvent Estimates in Dimension $d\geq 3$]
\label{lap}Let $d\geq 3$ and ${\mathfrak{e}}\in C^{3}(\Gamma ^{\ast };{%
\mathbb{R}})$ be a dispersion relation satisfying (\textbf{M}).

\begin{enumerate}
\item[(i)] If $c_{\ref{resolv}}\,\Phi _{2,2}(V)<1$ then there exists a
constant $c_{\ref{lap}.i}<\infty $ such that, for all $z\in \mathbb{C}%
\backslash \mathbb{R}$ and all $x,y\in \Gamma $, 
\begin{equation*}
|\langle \delta _{x}|(z-H(\,{\mathfrak{e,}}V))^{-1}\delta _{y}\rangle |\leq
c_{\ref{lap}.i}(1+|x|)^{2}(1+|y|)^{2}\text{ }. 
\end{equation*}

\item[(ii)] If $2\,c_{\ref{resolv}}\,\Phi _{2,3}(V)<1$ then there exists a
constant $c_{\ref{lap}.ii}<\infty $ such that, for all $z\in \mathbb{C}%
\backslash \mathbb{R}$ and all $x\in \Gamma $, 
\begin{equation*}
|\langle \delta _{x}|(z-i[H({\mathfrak{e}},V),A])^{-1}\delta _{y}\rangle
|\leq c_{\ref{lap}.ii}(1+|x|)^{2}(1+|y|)^{2}\text{ }. 
\end{equation*}
\end{enumerate}
\end{theorem}

\noindent \textit{Proof.} For $n\in {\mathbb{N}}$ and $z\in {\mathbb{C}}%
\backslash \mathbb{R}$ let%
\begin{equation*}
O_{n}(z)\doteq \lbrack V(z-h(\,{\mathfrak{e}})\,)^{-1}]^{n}=V^{\frac{1}{2}}%
\widetilde{O}_{n}(z)V^{\frac{1}{2}}(z-h(\,{\mathfrak{e}}))^{-1}\text{ }, 
\end{equation*}%
where 
\begin{equation*}
\widetilde{O}_{n}(z)\doteq \left[ V^{\frac{1}{2}}(z-h(\,{\mathfrak{e}}%
))^{-1}V^{\frac{1}{2}}\right] ^{n-1}\text{ }. 
\end{equation*}%
Assume that $c_{\ref{resolv}}\,\Phi _{2,2}(V)<1$. Then, by Lemma$\ $\ref%
{resolv}, $\Vert \widetilde{O}_{n}(z)\Vert <a^{n-1}$, for some $0<a<1$. Thus
we can define the operators 
\begin{equation*}
\widetilde{O}(z)\ \doteq \ \sum\limits_{n=1}^{\infty }\widetilde{O}_{n}(z)%
\text{ }. 
\end{equation*}%
It follows that, for each $z\in {\mathbb{C}}\backslash \mathbb{R}$, $\Vert 
\tilde{O}(z)\Vert _{\mathcal{B}[\ell ^{2}(\Gamma )]}\leq (1-a)^{-1}$ and
that $(z-h(\,{\mathfrak{e}})-V)$ has a bounded inverse given by 
\begin{equation*}
(z-h(\,{\mathfrak{e}})-V)^{-1}=(z-h(\,{\mathfrak{e}}))^{-1}\left[ 1+V^{\frac{%
1}{2}}\widetilde{O}(z)V^{\frac{1}{2}}(z-h(\,{\mathfrak{e}}))^{-1}\right] 
\text{ }. 
\end{equation*}%
This along with Lemma \ref{resolv} imply (i).

To prove (ii), we temporarily ignore questions of convergence and write 
\begin{equation*}
\left( z-i[H({\mathfrak{e}},V),A]\right) ^{-1}=\sum\limits_{n=0}^{\infty
}R_{0}(i[V,A]R_{0})^{n}\text{ }, 
\end{equation*}%
where $R_{0}\doteq \left( z-i[h(\,{\mathfrak{e}}),A]\right) ^{-1}$. Observe
that%
\begin{eqnarray*}
i[V,A] &=&iV^{\frac{1}{2}}(V^{\frac{1}{2}}A)+i(-AV^{\frac{1}{2}})V^{\frac{1}{%
2}} \\
&=&i\sum\limits_{\sigma =0}^{1}(-1)^{\sigma }(A^{\sigma }V^{\frac{1}{2}})(V^{%
\frac{1}{2}}A^{1-\sigma })\text{ }.
\end{eqnarray*}%
Hence,%
\begin{eqnarray*}
&&\left( z-i[H({\mathfrak{e}},V),A]\right) ^{-1}-R_{0} \\
&=&\sum\limits_{n=1}^{\infty }\sum\limits_{\sigma _{1},\ldots ,\sigma
_{n}=0}^{1}i^{n}(-1)^{|\underline{\sigma }|}R_{0}A^{\sigma _{1}}V^{\frac{1}{2%
}}\left( \prod\limits_{j=1}^{n-1}V^{\frac{1}{2}}A^{1-\sigma
_{j}}R_{0}A^{\sigma _{j+1}}V^{\frac{1}{2}}\right) V^{\frac{1}{2}}A^{1-\sigma
_{n}}R_{0}\text{ }.
\end{eqnarray*}%
Now, due to Lemma \ref{resolv}, we have that%
\begin{eqnarray*}
\left\Vert V^{\frac{1}{2}}A^{\sigma }R_{0}\delta _{x}\right\Vert _{2} &\leq
&c_{\ref{resolv}}(1+|x|^{2})\max \{\Phi _{2,2}(V)^{\frac{1}{2}},\Phi
_{2,3}(V)^{\frac{1}{2}}\} \\
&=&c_{\ref{resolv}}(1+|x|^{2})\Phi _{2,3}(V)^{\frac{1}{2}}\text{ }, \\
&& \\
\left\Vert V^{\frac{1}{2}}A^{\sigma }R_{0}A^{\eta }V^{\frac{1}{2}%
}\right\Vert _{\mathcal{B}[\ell ^{2}(\Gamma )]} &\leq &c_{\ref{resolv}}\max
\{\Phi _{2,2}(V),\Phi _{2,3}(V)\} \\
&=&c_{\ref{resolv}}\Phi _{2,3}(V)\text{ }.
\end{eqnarray*}%
for all $x\in \Gamma $ and $\sigma ,\eta \in \{0,1\}$. By assumption, $2c_{%
\ref{resolv}}\Phi _{2,3}(V)<1$, and the Neumann series evaluated on the
vectors $\delta _{x}$ and $\delta _{y}$, converges. Namely, 
\begin{eqnarray*}
|\langle \delta _{x}|(z-i[H({\mathfrak{e}},V),A])^{-1}\delta _{y}\rangle |
&\leq &c_{\ref{resolv}}^{2}(1+|x|^{2})(1+|y|^{2})\sum\limits_{n=0}^{\infty
}(2c_{\ref{resolv}}\Phi _{2,3}(V))^{n} \\
&=&\frac{c_{\ref{resolv}}^{2}(1+|x|^{2})(1+|y|^{2})}{1-2c_{\ref{resolv}}\Phi
_{2,3}(V)}\text{ }.
\end{eqnarray*}%
\hfill\ \hfill\ $\ \hfill \square $

\begin{corollary}
\label{th.pur.cont} Let $d\geq 3$ and ${\mathfrak{e\in }}C^{3}(\Gamma ^{\ast
};{\mathbb{R}})$ be a dispersion relation satisfying (\textbf{M}).

\begin{enumerate}
\item[(i)] If $c_{\ref{resolv}}\,\Phi _{2,2}(V)<1$ then $H({\mathfrak{e}},V)$
has purely absolutely continuous spectrum and 
\begin{equation*}
\sigma _{\mathrm{ac}}(H({\mathfrak{e}},V))=[0,{\mathfrak{e}}_{\max }]\text{ }%
. 
\end{equation*}

\item[(ii)] If $2\,c_{\ref{resolv}}\,\Phi _{2,3}(V)<1$ then $i[H({\mathfrak{e%
}},V),A]$ is positive and has purely absolutely continuous spectrum.
\end{enumerate}
\end{corollary}

\noindent \textit{Proof.} Assume that $c_{\ref{resolv}}\,\Phi _{2,2}(V)<1$.
From Theorem \ref{lap}$(i)$, for all $z\in {\mathbb{C}}\backslash {\mathbb{R}%
}$ and all vectors $\psi \in \mathrm{span}\{\delta _{x}\,|\,x\in \Gamma \}$,
i.e. $\psi $ of finite support, we have that 
\begin{equation*}
|\langle \psi |(z-H(\,{\mathfrak{e,}}V))^{-1}\psi \rangle |\leq c(\psi
)<\infty 
\end{equation*}%
with $c(\psi )$ depending only on $\psi $. As $\mathrm{span}%
\{\delta_{x}\,|\,x\in \Gamma \}$ is dense in $\ell ^{2}(\Gamma )$, this last
estimate implies the absolute continuity of the spectrum of $H(\mathfrak{e}%
,V)$. See, for instance, \cite[Proposition 4.1]{Cycon}. Analogously, by
Theorem \ref{lap} (ii), $i[H({\mathfrak{e}},V),A]$ has only absolutely
continuous spectrum whenever $2\,c_{\ref{resolv}}\Phi _{2,3}(V)<1$. If $\Phi
_{2,2}(V)<\infty $ then $V$ and $i[V,A]$ define trace class operators. By
the Kato-Rosenblum theorem, 
\begin{eqnarray*}
\sigma _{\mathrm{ac}}(H({\mathfrak{e}},V)) &=&\sigma _{\mathrm{ac}}(\,h({%
\mathfrak{e)}}\,)=[0,{\mathfrak{e}}_{\mathrm{\max }}]\text{ }, \\
\sigma _{\mathrm{ac}}(i[H({\mathfrak{e}},V),A]) &=&\sigma _{\mathrm{ac}}(i[h(%
{\mathfrak{e)}},A])\subset \mathbb{R}_{0}^{+}\text{ }.
\end{eqnarray*}

\hfill $\square $

\section{Bound on $N_{\mathrm{pp}}[\mathfrak{e},V]$}

\label{sect.size.pp}

The positivity of the commutator $i[H({\mathfrak{e}},V^{\prime }),A]$ at
small $\Phi _{2,3}(V^{\prime })$, stated\ in Corollary \ref{th.pur.cont},
yields upper bounds on $N_{\mathrm{pp}}[\mathfrak{e},V]$ for $d\geq 3$,
without assuming that $V$ has a finite support:

\begin{corollary}[{Upper Bound on $N_{pp}[\mathfrak{e},V]$, Infinite Range
Case, $d\geq 3$}]
\label{koro.u.b.N.2} Let $d\geq 3$ and ${\mathfrak{e}}$ be a dispersion
relation from $C^{4}(\Gamma ^{\ast };{\mathbb{R}})$ satisfying (\textbf{M}).
Let $V$ be a potential with $\Phi _{2,3}(V)<\infty $ and choose $V_{1},V_{2}$
such that $2\,c_{\ref{resolv}}\,\Phi _{2,3}(V_{1})<1$ and $V_{2}$ has finite
support. Then 
\begin{equation*}
N_{\mathrm{pp}}[\mathfrak{e},V]\leq \mathrm{Tr}\left[ E_{-}(i[V_{2},A])%
\right] \text{ }. 
\end{equation*}
\end{corollary}

\noindent \textit{Proof.} If $2\,c_{\ref{resolv}}\,\Phi _{2,3}(V_{1})<1$
then, by Corollary \ref{th.pur.cont}, $i[H({\mathfrak{e}},V_{1}),A]\geq 0$
and has purely absolutely continuous spectrum. Thus, by Lemma \ref{Virial},
if $\psi $ is an eigenvector of $H({\mathfrak{e}},V)$ then $\langle \psi
\,|\,i[V_{2},A]\psi \rangle <0$ . Ergo, $N_{\mathrm{pp}}[\mathfrak{e},V]\leq 
\mathrm{Tr}\left[ E_{-}(i[V_{2},A])\right] $. See the proof of Corollary \ref%
{koro.fin.ren.u.b} for more details. \hfill $\square $\smallskip

From the corollary, 
\begin{equation}
N_{\mathrm{pp}}[\mathfrak{e},V]\leq \min \left\{ \mathrm{Tr}\left[
E_{-}(i[V^{\prime },A])\right] \text{ }|\text{ }|\mathrm{supp}\text{ }%
V^{\prime }|<\infty \text{ },\text{ }\Phi _{2,3}(V-V^{\prime })<\frac{c_{\ref%
{resolv}}^{-1}}{2}\right\} \text{ }.  \label{main bound}
\end{equation}%
Recall that $E_{-}(i[V^{\prime },A])$ is the spectral projector associated
with the negative spectrum of the (selfadjoint) commutator $i[V^{\prime },A]$%
. As the operator $h(\mathfrak{e})$ (the hopping matrix the Schr\"{o}dinger
operator $H({\mathfrak{e}},V)$) is invariant with respect to translations,
for all $z\in \mathbb{Z}^{d}$, 
\begin{equation*}
N_{\mathrm{pp}}[\mathfrak{e},V]=N_{\mathrm{pp}}[\mathfrak{e},V^{(z)}]\text{ }%
, 
\end{equation*}%
where $V^{(z)}$ is the translation (\ref{transla V}) of the potential $V$.
From this remark, Corollary \ref{koro.fin.ren.u.b copy(1)} and the estimate (%
\ref{main bound}), we arrive at our main result:

\begin{theorem}[{Bound on $N_{\mathrm{pp}}[\mathfrak{e},V]$, $d\geq 3$,
infinite range case}]
\label{th main}Let $d\geq 3$ and ${\mathfrak{e}}$ be a dispersion relation
from $C^{4}(\Gamma ^{\ast };{\mathbb{R}})$ satisfying (\textbf{M}). Then,
for the finite constant $c_{\ref{th main}}\doteq \left( 2\,c_{\ref{resolv}%
}\right) ^{-1}$, 
\begin{equation*}
N_{\mathrm{pp}}[\mathfrak{e},V]\leq \inf \left\{ |\mathrm{supp}\text{ }%
V^{\prime }|\text{ }|\text{ }z\in \mathbb{Z}^{d}\text{ },\text{ }V^{\prime }%
\text{ with }\Phi _{2,3}(V^{(z)}-V^{\prime })<c_{\ref{th main}}\right\} 
\text{ }. 
\end{equation*}
\end{theorem}

The above estimate on $N_{\mathrm{pp}}[\mathfrak{e},V]$ implies the absence
of embedded eigenvalues of $H({\mathfrak{e}},V)$ for a class of potentials $%
V $:

\begin{corollary}[Absence of embedded eigenvalues]
\label{koro.embedded.ev} Let $d\geq 3$ and ${\mathfrak{e}}$ be a dispersion
relation from $C^{4}(\Gamma ^{\ast };{\mathbb{R}})$ satisfying (\textbf{M}).
Let $V$ be a potential with $\Phi _{2,3}(V)<\infty $. Assume that $%
V=V_{1}+V_{2}$, where $\Phi _{2,3}(V_{2}^{(z)})<c_{\ref{th main}}$ for some
translation $z\in \mathbb{Z}^{d}$, $V_{1}$ has a finite support and $H({%
\mathfrak{e}},V)$ has exactly $|\mathrm{supp}$ $V_{1}|$ discrete
eigenvalues, counting their multiplicities. Then all eigenvalues of $H({%
\mathfrak{e}},V)$ are discrete.
\end{corollary}

\section{Appendix\label{Appendix}}

\subsection{Proof of Lemma \protect\ref{Mourre}\label{proof.moure}}

Let $N$ be the unique self-adjoint extension of the operator $\tilde{N}$
defined on $C^{\infty }(\Gamma ^{\ast };{\mathbb{C}})\subset L^{2}(\Gamma
^{\ast })$ by 
\begin{equation*}
\tilde{N}{\hat{\psi}}(p)=\sum\limits_{i=1}^{d}(1-\partial _{p_{i}}^{2}){\hat{%
\psi}}(p). 
\end{equation*}%
Observe that for some $\mathrm{const}<\infty $ and all ${\hat{\psi}},{\hat{%
\psi}}^{\prime }\in C^{\infty }(\Gamma ^{\ast };{\mathbb{C}})$, 
\begin{equation*}
|\langle {\hat{\psi}}^{\prime }\,|\,A{\hat{\psi}}\rangle |\leq \mathrm{const}%
\,\parallel {\hat{\psi}}_{2}^{\prime }\parallel _{2}\parallel N^{\frac{1}{2}}%
{\hat{\psi}}_{2}\parallel _{2}\leq \mathrm{const}\,\parallel N^{\frac{1}{2}}{%
\hat{\psi}}^{\prime }\parallel _{2}\parallel N^{\frac{1}{2}}{\hat{\psi}}%
_{2}\parallel _{2}. 
\end{equation*}%
For all ${\hat{\psi}}\in C^{\infty }(\Gamma ^{\ast };{\mathbb{C}})$, 
\begin{eqnarray*}
(NA-AN){\hat{\psi}}(p) &=&-i\sum\limits_{k,k^{\prime }=1}^{d}\left\{
(2[\partial _{p_{k}}^{2}\partial _{p_{k^{\prime }}}{\mathfrak{e}}%
(p)][\partial _{p_{k^{\prime }}}{\hat{\psi}}(p)]+\frac{1}{2}[\partial
_{p_{k}}^{2}\partial _{p_{k^{\prime }}}^{2}{\mathfrak{e}}(p)]{\hat{\psi}}%
(p)\right. \\
&&\left. +2[\partial _{p_{k}}\partial _{p_{k^{\prime }}}{\mathfrak{e}}%
(p)][\partial _{p_{k}}\partial _{p_{k^{\prime }}}{\hat{\psi}}(p)]\right\} 
\text{ }.
\end{eqnarray*}%
An integration of the terms with second derivatives of ${\hat{\psi}}$ by
parts yields, for some $0<\mathrm{const}<\infty $ and all ${\hat{\psi}},{%
\hat{\psi}}^{\prime }\in C^{\infty }(\Gamma ^{\ast };{\mathbb{C}})$, that%
\begin{equation*}
\left\vert \langle N{\hat{\psi}}^{\prime }\,|\,A{\hat{\psi}}\rangle -\langle
A{\hat{\psi}}^{\prime }\,|\,N{\hat{\psi}}\rangle \right\vert \text{ }\leq 
\text{ }\mathrm{const}\parallel N^{\frac{1}{2}}{\hat{\psi}}_{2}^{\prime
}\parallel _{2}\parallel N^{\frac{1}{2}}{\hat{\psi}}_{2}\parallel _{2}\text{ 
}. 
\end{equation*}%
Thus, by Nelson's commutator theorem (see \cite[ Theorem X.36]{RS3}), $A$ is
essentially self-adjoint on $C^{\infty }(\Gamma ^{\ast };{\mathbb{C}})$.

Clearly, as $\chi $ is continuous and has compact support, 
\begin{eqnarray*}
\lefteqn{\chi (H(\mathfrak{e},V))-\chi (H(\mathfrak{e},0))} \\
&=&\lim\limits_{\eta \downarrow 0}\frac{1}{\sqrt{\pi \eta }}%
\int\nolimits_{0}^{\infty }\,\chi (t)\left[ \exp \left( \frac{-(H({\mathfrak{%
e}},V)-t)^{2}}{\eta }\right) -\exp \left( \frac{-(H({\mathfrak{e}},0)-t)^{2}%
}{\eta }\right) \right] \mathrm{d}t
\end{eqnarray*}%
in norm sense. Observe that%
\begin{eqnarray}
\lefteqn{-\eta \int\nolimits_{0}^{\infty }\chi (t)\left[ \exp \left( \frac{%
-(H(\mathfrak{e},V)-t)^{2}}{\eta }\right) -\exp \left( \frac{-(H(\mathfrak{e}%
,0)-t)^{2}}{\eta }\right) \right] \mathrm{d}t}  \label{mourre.eq.2} \\
&=&\int\nolimits_{0}^{1}\left[ \int\nolimits_{0}^{\infty }\,\chi (t)\exp
\left( \frac{-s(H({\mathfrak{e}},V)-t)^{2}}{\eta }\right) (Vh({\mathfrak{e}}%
)+h({\mathfrak{e}})V+V^{2}-2tV)\right.  \notag \\
&&\left. \exp \left( \frac{-(1-s)(H({\mathfrak{e}},0)-t)^{2}}{\eta }\right) 
\mathrm{d}t\right] \mathrm{d}s\text{ }.  \notag
\end{eqnarray}%
As $V$ is a compact operator, it follows from (\ref{mourre.eq.2}) that $\chi
(H({\mathfrak{e}},V))-\chi (H({\mathfrak{e}},0))$ is compact.

The difference 
\begin{equation*}
i[H({\mathfrak{e}},V),A]-i[H({\mathfrak{e}},0),A]=i[V,A]
\end{equation*}%
is also a compact operator, by assumption. To finish the proof observe that $%
i[H({\mathfrak{e}},0),A]$ is unitarily equivalent to the multiplication
operator $|\nabla {\mathfrak{e}}|^{2}$. Moreover, 
\begin{equation*}
|\nabla {\mathfrak{e}}|^{2}\cdot \chi ^{2}(H({\mathfrak{e}},0))\geq c\cdot
\chi ^{2}(H({\mathfrak{e}},0))
\end{equation*}
is bounded below on the range of $\chi ^{2}(H({\mathfrak{e}},0))$ by a
positive multiple of the identity, since $\chi ^{2}(H({\mathfrak{e}},0))$ is
supported away from the thresholds. Thus, there is a constant $c_{\chi
}^{0}>0$ such that 
\begin{equation*}
\chi (H({\mathfrak{e}},0))i[H({\mathfrak{e}},0),A]\chi (H({\mathfrak{e}}%
,0))\geq c_{\chi }^{0}\chi ^{2}(H({\mathfrak{e}},0)).
\end{equation*}
$\ \hfill \square $

\subsection{Proof of Lemma \protect\ref{Virial}\label{proof.virial}}

Let $\psi $ be an eigenvector of $H({\mathfrak{e}},V_{1}+V_{2})$ and define,
for each $n\in \mathbb{Z}\backslash \{0\}$, the vector 
\begin{equation*}
\psi _{n}\doteq \frac{i\,n}{i\,n+A}\psi \text{ }. 
\end{equation*}%
Since $i[H({\mathfrak{e}},V_{1}),A]$ and $i[V_{2},A]$ are bounded operators,
by assumption, we have that%
\begin{eqnarray*}
\lim_{n\rightarrow \infty }\langle \psi _{-n}\,|\,i[H({\mathfrak{e}}%
,V_{1}),A]\psi _{n}\rangle &=&\langle \psi \,|\,i[H({\mathfrak{e}}%
,V_{1}),A]\psi \rangle , \\
\lim_{n\rightarrow \infty }\langle \psi _{-n}\,|\,i[V_{2},A]\psi _{n}\rangle
&=&\langle \psi \,|\,i[V_{2},A]\psi \rangle \text{ }.
\end{eqnarray*}%
Note that 
\begin{eqnarray*}
\lefteqn{\langle \psi _{-n}\,|\,i[H({\mathfrak{e}},V_{1}+V_{2}),A]\psi
_{n}\rangle } \\
&=&\langle \psi _{-n}\,|\,i[H({\mathfrak{e}},V_{1}),A]\psi _{n}\rangle
+\langle \psi _{-n}\,|\,i[V_{2},A]\psi _{n}\rangle \text{ }.
\end{eqnarray*}%
Hence it suffices to prove, for all $n\in {\mathbb{N}}$, that%
\begin{equation*}
\langle \psi _{-n}\,|\,i[H({\mathfrak{e}},V_{1}+V_{2}),A]\psi _{n}\rangle =0%
\text{ }. 
\end{equation*}%
This is easily seen, however, as for all $n\in {\mathbb{N}}$, 
\begin{eqnarray*}
\lefteqn{\langle \psi _{-n}\,|\,i[H({\mathfrak{e}},V_{1}+V_{2}),A]\psi
_{n}\rangle } \\
&=&\left\langle \psi \,\Big|\,i\left[ H({\mathfrak{e}},V_{1}+V_{2}),\frac{%
i\,nA}{i\,n+A}\right] \psi \right\rangle =0\text{ }.
\end{eqnarray*}

\hfill $\square $

\subsection{Proof of Lemma \protect\ref{resolv}\label{proof.resolv}}

In order to prove Lemma \ref{resolv} we need the following estimate:

\begin{lemma}
\label{estimates.Morse.func} Assume that $d\geq 3$ and let ${\mathfrak{e}}$
be a dispersion relation with $K({\mathfrak{e}})>0$ and $\left\Vert {%
\mathfrak{e}}\right\Vert _{C^{3}}<\infty $. Suppose that $\chi \in
C^{2}(\Gamma ^{\ast };{\mathbb{R}})$. Then there exists a constant $\mathrm{c%
}_{\ref{estimates.Morse.func}}<\infty $ depending only on $K({\mathfrak{e}})$%
, $\left\Vert {\mathfrak{e}}\right\Vert _{C^{3}}$ and $\left\Vert \chi
\right\Vert _{C^{2}}$ such that 
\begin{equation*}
\left\vert \int_{\Gamma ^{\ast }}\frac{\chi (p)}{z-{\mathfrak{e}}(p)}\mathrm{%
d}\mu ^{\ast }(p)\right\vert \leq \mathrm{c}_{\ref{estimates.Morse.func}} 
\text{,} 
\end{equation*}%
for all $z\in {\mathbb{C}}\backslash {\mathbb{R}}$.
\end{lemma}

\noindent \textit{Proof.} We assume w.l.o.g. that $z$ is bounded by $|z|\leq 
{\mathfrak{e}}_{\max }+1$, say. We further note that ${\mathfrak{e}}$ has
only finitely many critical points, $\#Q<\infty $, abbreviating $Q\doteq 
\mathrm{Crit}({\mathfrak{e}})$, since $\Gamma ^{\ast }$ is compact and ${%
\mathfrak{e}}$ is a Morse function. The latter is also the reason that, for
each $q\in Q$, there exist an index $m_{q}\in \{0,\ldots ,d\}$ and a $C^{2}$%
--coordinate chart $\xi _{q}\in C^{2}(B_{d-m}\times B_{m};\mathcal{U}_{q})$,
for%
\begin{equation*}
B_{n}\doteq B_{\mathbb{R}^{n}}(0,r)=\{x\in \mathbb{R}^{n}\text{ }:\text{ }%
|x|<r\},\;r>0\text{ }, 
\end{equation*}%
denoting the Euclidean open ball in $\mathbb{R}^{n}$ of radius $r$ and $%
\mathcal{U}_{q}\subset \Gamma ^{\ast }$ being an open neighborhood of $q$
such that, for all $x\in B_{d-m}$, $y\in B_{m}$,%
\begin{equation*}
c_{1}\leq |\det \mathrm{Jac}\text{ }\xi _{q}(x,y)|\leq c_{2}\text{ }, 
\end{equation*}%
\begin{equation*}
{\mathfrak{e\circ }}\xi _{q}(x,y)={\mathfrak{e}}(q)+x^{2}-y^{2}\text{ }, 
\end{equation*}%
\begin{equation*}
\mathcal{U}_{q}\supseteq B_{\Gamma ^{\ast }}(q,\delta )\text{ }, 
\end{equation*}%
for suitable constants $c_{1},\delta >0$, $r\in (0,1)$ and $c_{2}<\infty $. $%
\delta >0$ can be chosen such that away from the critical points we can find
a finite set%
\begin{equation*}
\widetilde{Q}\subseteq \mathcal{N}\doteq \{q\in \Gamma ^{\ast }\text{ }|%
\text{ }{\mathfrak{e}}(q)=\mathrm{Re}\{z\}\} 
\end{equation*}%
and, for each $q\in \widetilde{Q}$, a $C^{2}$--coordinate chart $\tilde{\xi}%
_{q}\in C^{2}((-r,r)\times B_{d-1};\widetilde{\mathcal{U}}_{q})$, \ with $%
\widetilde{\mathcal{U}}_{q}\subset \Gamma ^{\ast }$ being an open
neighborhood of $q$, such that, for all $x\in $ $(-r,r)$, $y\in B_{d-1}$,%
\begin{equation*}
c_{1}\leq |\det \mathrm{Jac}\text{ }\tilde{\xi}_{q}(x,y)|\leq c_{2}\text{ }, 
\end{equation*}%
\begin{equation*}
{\mathfrak{e\circ }}\tilde{\xi}_{q}(x,y)={\mathfrak{e}}(q)+x\text{ }, 
\end{equation*}%
\begin{equation*}
\bigcup\limits_{q\in \widetilde{Q}}\widetilde{\mathcal{U}}_{q}\supseteq
\{p\in \Gamma ^{\ast }\text{ }:\text{ }|{\mathfrak{e}}(q)-z|<\delta ,\text{
dist}(p,Q)\geq \delta \}\text{ }. 
\end{equation*}%
Let 
\begin{equation*}
\widehat{\mathcal{N}}\doteq \left\{ p\in \Gamma ^{\ast }\text{ }:\text{ }|{%
\mathfrak{e}}(p)-z|>\frac{\delta }{2}\right\} \text{ }. 
\end{equation*}%
Then $\{\widehat{\mathcal{N}}\}\cup \{\mathcal{U}_{q}\}_{q\in Q}\cup \{%
\widetilde{\mathcal{U}}_{q}\}_{q\in \widetilde{Q}}$ is a finite open
covering of $\Gamma ^{\ast }$ and there exists a subordinate partition of
unity, 
\begin{equation*}
\{\hat{\eta}\}\cup \{\eta _{q}\}_{q\in Q}\cup \{\widetilde{\eta }%
_{q}\}_{q\in \widetilde{Q}}\subseteq C^{\infty }(\Gamma ^{\ast };[0,1])\text{
}, 
\end{equation*}
such that 
\begin{equation}
\mathrm{supp}\text{ }\hat{\eta}\subset \widehat{\mathcal{N}}, \qquad \mathrm{%
supp}\text{ }\eta _{q}\subset \mathcal{U}_{q}, \qquad \mathrm{supp}\; 
\widetilde{\eta }_{q} \subset \widetilde{\mathcal{U}}_{q}\text{ },
\label{supp incl}
\end{equation}
for $q\in Q\cup \widetilde{Q}$, and 
\begin{equation*}
\hat{\eta}+\sum\limits_{q\in Q}\eta _{q}+\sum\limits_{q\in \widetilde{Q}}%
\widetilde{\eta }_{q}\equiv 1\text{ }. 
\end{equation*}%
It follows that%
\begin{equation*}
\int_{\Gamma ^{\ast }}\frac{\chi (p)}{z-{\mathfrak{e}}(p)}\mathrm{d}\mu
^{\ast }(p)=\widehat{\mathcal{I}}+\sum\limits_{q\in Q}\mathcal{I}%
_{q}+\sum\limits_{q\in \widetilde{Q}}\widetilde{\mathcal{I}}_{q}\text{ }, 
\end{equation*}%
where%
\begin{eqnarray*}
\widehat{\mathcal{I}} &\doteq &\int_{\Gamma ^{\ast }}\frac{\hat{\eta}(p)\chi
(p)}{z-{\mathfrak{e}}(p)}\mathrm{d}\mu ^{\ast }(p)\text{ }, \\
\widetilde{\mathcal{I}}_{q} &\doteq &\int_{B_{d-1}}\mathrm{d}%
^{d-1}y\int_{-r}^{r}\mathrm{d}x\frac{\widetilde{f}_{q}(x,y)}{ib-x}\text{ },
\\
\mathcal{I}_{q} &\doteq &\int_{B_{d-m_{q}}}\mathrm{d}^{d-m_{q}}x%
\int_{B_{m_{q}}}\mathrm{d}^{m_{q}}y\frac{f_{q}(x,y)}{a_{q}+ib-x^{2}+y^{2}}%
\text{ },
\end{eqnarray*}%
where $b\doteq \mathrm{Im}\{z\}$, $a_{q}\doteq \mathrm{Re}\{z\}-{\mathfrak{e}%
}(q) $, and%
\begin{eqnarray*}
\widetilde{f}_{q} &\doteq &(\widetilde{\eta }_{q}\circ \tilde{\xi}_{q})(\chi
\circ \tilde{\xi}_{q})|\det \mathrm{Jac}\text{ }\tilde{\xi}_{q}|\text{ }, \\
f_{q} &\doteq &(\eta _{q}\circ \xi _{q})(\chi \circ \xi _{q})|\det \mathrm{%
Jac}\text{ }\xi _{q}|\text{ }.
\end{eqnarray*}%
Note that $\widetilde{f}_{q}\in C_{0}^{2}((-r,r)\times B_{d-1};\mathbb{R})$
and $f_{q}\in C_{0}^{2}(B_{d-m_{q}}\times B_{m_{q}};\mathbb{R})$, due to (%
\ref{supp incl}). Moreover, $\Vert \widetilde{f}_{q}\Vert
_{C^{2}},\left\Vert f_{q}\right\Vert _{C^{2}}<\infty $. The asserted
estimate now follows from Lemmata \ref{Lemma1}--\ref{Lemma3} and the trivial
estimate%
\begin{equation*}
|\widehat{\mathcal{I}}|\leq \frac{2}{\delta }\int_{\Gamma ^{\ast }}|\chi (p)|%
\mathrm{d}\mu ^{\ast }(p)\text{ }. 
\end{equation*}%
Observe that the constants $r,\delta ,c_{1},c_{2}$ and $\#Q$, $\#\widetilde{Q%
}$ only depend on $K({\mathfrak{e}})$ , $\left\Vert {\mathfrak{e}}%
\right\Vert _{C^{3}}$ and $\left\Vert \chi \right\Vert _{C^{2}}$. \hfill $%
\square $

\begin{lemma}
\label{Lemma1}Assume that $d\geq 1$ and $0<r<1$. There is a constant $%
\widehat{C}_{1}<\infty $ such that, for all $f\in C^{1}((-r,r)\times B_{d-1};%
\mathbb{R})$ and all $b\in \mathbb{R}\backslash \{0\}$,%
\begin{equation*}
\left\vert \int_{B_{d-1}}\mathrm{d}^{d-1}y\int_{-r}^{r}\mathrm{d}x\frac{%
f(x,y)}{ib-x}\right\vert \leq \widehat{C}_{1}\left\Vert f\right\Vert _{C^{1}}%
\text{ }. 
\end{equation*}
\end{lemma}

\noindent \textit{Proof. }For all $x\in (-r,r)$ and all $y\in B_{d-1}$, the
fundamental theorem of calculus gives%
\begin{equation*}
\left\vert \frac{f(x,y)-f(0,y)}{ib-x}\right\vert \leq \left\vert \frac{x}{%
ib-x}\right\vert \left\Vert \partial _{x}f\right\Vert _{\infty }\leq
\left\Vert f\right\Vert _{C^{1}}\text{ }, 
\end{equation*}%
and thus%
\begin{equation*}
\left\vert \int_{B_{d-1}}\mathrm{d}^{d-1}y\int_{-r}^{r}\mathrm{d}x\frac{%
f(x,y)}{ib-x}\right\vert \leq 2|B_{d-1}|\left\Vert f\right\Vert
_{C^{1}}\left( 1+\left\vert \int_{-r}^{r}\frac{\mathrm{d}x}{ib-x}\right\vert
\right) \text{ }. 
\end{equation*}%
The assertion follows then from%
\begin{equation*}
\left\vert \int_{-r}^{r}\frac{\mathrm{d}x}{ib-x}\right\vert \leq \left\vert
\int_{-1}^{1}\frac{b\text{ }\mathrm{d}x}{b^{2}+x^{2}}\right\vert =2\arctan
(|b|)\leq \pi \text{ }. 
\end{equation*}

\hfill $\square $

\begin{lemma}
\label{Lemma2}Assume that $d\geq 3$ and $0<r<1$. There is a constant $%
\widehat{C}_{2}<\infty $ such that, for all $f\in C_{0}^{1}(B_{d};\mathbb{R}%
) $, all $a\in \mathbb{R}$ and all $b\in \mathbb{R}\backslash \{0\}$,%
\begin{equation*}
\left\vert \int_{B_{d}}\frac{f(x)}{a+ib-x^{2}}\mathrm{d}^{d}x\right\vert
\leq \widehat{C}_{2}\left\Vert f\right\Vert _{C^{1}}(1+a^{2}+b^{2})\text{ }. 
\end{equation*}
\end{lemma}

\noindent \textit{Proof.} Introducing spherical coordinates, we observe that 
\begin{equation*}
\mathcal{J}\doteq \int_{B_{d}}\frac{f(x)}{a+ib-x^{2}}\mathrm{d}%
^{d}x=\int_{0}^{r}\frac{g(s)s^{d-1}}{a+ib-s^{2}}\mathrm{d}s, 
\end{equation*}%
where $g\in C^{1}([0,1];\mathbb{R})$ is the spherical average of $f$, $%
\left\Vert g\right\Vert _{C^{1}}\leq \left\Vert f\right\Vert _{C^{1}}$,
defined by%
\begin{equation*}
g(s)\doteq \int_{\mathbb{S}^{d-1}}f(s\vartheta )\,\mathrm{d}^{d-1}\sigma
(\vartheta )\text{ }. 
\end{equation*}%
An integration by parts gives%
\begin{eqnarray*}
\mathrm{Re}\{\mathcal{J}\} &=&\int_{0}^{r}g(s)s^{d-1}\frac{a-s^{2}}{%
(a-s^{2})^{2}+b^{2}}\mathrm{d}s \\
&=&-\frac{1}{4}\int_{0}^{r}g(s)s^{d-2}\left( \frac{\mathrm{d}}{\mathrm{d}s}%
\ln \left[ (a-s^{2})^{2}+b^{2}\right] \right) \mathrm{d}s \\
&=&\frac{1}{4}\int_{0}^{r}\left( g^{\prime
}(s)s^{d-2}+(d-2)g(s)s^{d-3}\right) \ln \left[ (a-s^{2})^{2}+b^{2}\right] 
\mathrm{d}s\text{ }.
\end{eqnarray*}%
We used above that $g(r)=0$. Now, use the elementary estimate%
\begin{equation}
|\ln (\lambda )|\leq \frac{1}{2\alpha }(\lambda ^{\alpha }+\lambda ^{-\alpha
})\text{ },  \label{28}
\end{equation}%
which holds true for all $\lambda ,\alpha >0$. Choosing $\alpha \doteq \frac{%
1}{8}$, (\ref{28}) yields%
\begin{eqnarray}
\left\vert \mathrm{Re}\{\mathcal{J}\}\right\vert &\leq &\frac{d-1}{4}%
\left\Vert g\right\Vert _{C^{1}}\int_{0}^{1}\left\vert \ln \left[
(a-s^{2})^{2}+b^{2}\right] \right\vert \mathrm{d}s  \notag \\
&\leq &C\left\Vert g\right\Vert _{C^{1}}\left[ (1+a^{2}+b^{2})^{\frac{1}{8}%
}+\int_{0}^{1}\frac{\mathrm{d}s}{|a-s^{2}|^{\frac{1}{4}}}\right]  \notag \\
&\leq &C^{\prime }\left\Vert g\right\Vert _{C^{1}}(1+a^{2}+b^{2})\text{ },
\end{eqnarray}%
for some universal constants $C,C^{\prime }<\infty $. Similarly,%
\begin{eqnarray*}
\left\vert \mathrm{Im}\{\mathcal{J}\}\right\vert &=&\frac{1}{|b|}\left\vert
\int_{0}^{r}\frac{g(s)s^{d-1}}{1+b^{-2}(a-s^{2})^{2}}\mathrm{d}s\right\vert
\\
&=&\frac{1}{2}\left\vert \int_{0}^{r}g(s)s^{d-2}\left( \frac{\mathrm{d}}{%
\mathrm{d}s}\arctan \left[ \frac{a-s^{2}}{|b|}\right] \right) \mathrm{d}%
s\right\vert \\
&=&\frac{1}{2}\left\vert \int_{0}^{r}\left( g^{\prime
}(s)s^{d-2}+(d-2)g(s)s^{d-3}\right) \arctan \left[ \frac{a-s^{2}}{|b|}\right]
\mathrm{d}s\right\vert \\
&\leq &C\left\Vert g\right\Vert _{C^{1}}.
\end{eqnarray*}

\hfill $\square $

\begin{lemma}
\label{Lemma3}Assume that $d\geq 3$, $0<r<1$, and $1\leq m\leq d-1$. There
is a constant $\widehat{C}_{3}<\infty $ such that, for all $f\in
C_{0}^{1}(B_{d-m}\times B_{m};\mathbb{R})$, all $a\in \mathbb{R}$ and all $%
b\in \mathbb{R}\backslash \{0\}$,%
\begin{equation*}
\left\vert \int_{B_{d-m}}\int_{B_{m}}\frac{f(x,y)}{a+ib-x^{2}+y^{2}}\mathrm{d%
}^{d-m}x\text{ }\mathrm{d}^{m}y\right\vert \leq \widehat{C}_{3}\left\Vert
f\right\Vert _{C^{1}}(1+a^{2}+b^{2})\text{ }. 
\end{equation*}
\end{lemma}

\noindent \textit{Proof. }As in Lemma \ref{Lemma2}, we introduce spherical
coordinates on $B_{d-m}$ and $B_{m}$ and define $g\in C^{1}([0,1]\times
\lbrack 0,1];\mathbb{R})$, with $\left\Vert g\right\Vert _{C^{1}}\leq
\left\Vert f\right\Vert _{C^{1}}$, by%
\begin{equation*}
g(x,y)\doteq \int_{\mathbb{S}^{d-m-1}}\int_{\mathbb{S}^{m-1}}f(x\vartheta
,y\kappa )\text{ }\mathrm{d}^{d-m-1}\sigma (\vartheta )\,\mathrm{d}%
^{m-1}\sigma (\kappa )\text{ }, 
\end{equation*}%
so that 
\begin{eqnarray*}
\mathcal{K} &\doteq &\int_{B_{d-m}}\int_{B_{m}}\frac{f(x,y)}{a+ib-x^{2}+y^{2}%
}\,\mathrm{d}^{d-m}x\text{ }\mathrm{d}^{m}y \\
&=&\int_{0}^{r}\int_{0}^{r}\frac{g(x,y)x^{d-m-1}y^{m-1}}{a+ib-x^{2}+y^{2}}\,%
\mathrm{d}x\text{ }\mathrm{d}y\text{ }.
\end{eqnarray*}%
We perform yet another smooth coordinate change by 
\begin{equation*}
\phi :(0,r)\times (-1,1)\rightarrow (0,2r)\times (0,2r)\text{ }, 
\end{equation*}
\begin{equation*}
x=\phi _{1}(s,u)\doteq s(1+u),\text{ \ \ }y=\phi _{2}(s,u)\doteq s(1-u)\text{
}. 
\end{equation*}%
With this definition, $\left\vert \det \mathrm{Jac}\text{ }\phi
(s,u)\right\vert =2s$ and we obtain 
\begin{equation*}
\mathcal{K}=2\int_{0}^{r}\int_{-1}^{1}\frac{s^{d-1}\widetilde{g}(s,u)}{%
a+ib-(2s)^{2}u}\mathrm{d}u\text{ }\mathrm{d}r\text{ }, 
\end{equation*}%
where%
\begin{equation*}
\widetilde{g}(s,u)\doteq \left( 1+u\right) ^{d-m-1}\left( 1-u\right)
^{m-1}g\left( s(1+u),s(1-u)\right) \text{ }. 
\end{equation*}
Note that $\widetilde{g}(s,u)=0$ whenever $s(1\pm u)\geq r$. Following a
similar strategy as in the proof of Lemma \ref{Lemma2}, we first observe
that 
\begin{eqnarray*}
\mathrm{Re}\{\mathcal{K}\} &=&2\int_{0}^{r}\int_{-1}^{1}s^{d-1}\widetilde{g}%
(s,u)\frac{a-(2s)^{2}u}{(a-(2s)^{2}u)^{2}+b^{2}}\text{ }\mathrm{d}u\text{ }%
\mathrm{d}s \\
&=&-\frac{1}{4}\int_{0}^{r}s^{d-3}\left[ \int_{-1}^{1}\widetilde{g}%
(s,u)\left( \frac{\mathrm{d}}{\mathrm{d}u}\ln \left[ (a-(2s)^{2}u)^{2}+b^{2}%
\right] \right) \text{ }\mathrm{d}u\right] \mathrm{d}s \\
&=&-\frac{1}{4}\int_{0}^{r}s^{d-3}\widetilde{g}(s,1)\ln \left[
(a-(2s)^{2})^{2}+b^{2}\right] \mathrm{d}s \\
&&+\frac{1}{4}\int_{0}^{r}s^{d-3}\widetilde{g}(s,-1)\ln \left[
(a+(2s)^{2})^{2}+b^{2}\right] \mathrm{d}s \\
&&+\frac{1}{4}\int_{0}^{r}s^{d-3}\left[ \int_{-1}^{1}(\partial _{u}%
\widetilde{g})(s,u)\ln \left[ (a-(2s)^{2}u)^{2}+b^{2}\right] \text{ }\mathrm{%
d}u\right] \mathrm{d}s\text{ }.
\end{eqnarray*}%
We use (\ref{28}) again to bound%
\begin{equation*}
\ln \left[ (a-(2s)^{2}u)^{2}+b^{2}\right] \leq 8(8+2a^{2}+b^{2})^{\frac{1}{8}%
}+8|(2s)^{2}|u|-|a||^{-\frac{1}{4}} 
\end{equation*}%
for $u=\pm 1$ and $u\in \lbrack -1,1]$, respectively, and hence%
\begin{eqnarray*}
\left\vert \mathrm{Re}\{\mathcal{K}\}\right\vert &\leq &C\left\Vert
f\right\Vert _{C^{1}}\Big(1+a^{2}+b^{2}+\int_{0}^{2}\frac{\mathrm{d}s}{%
|s^{2}-|a||^{\frac{1}{4}}} \\
&&+\int_{0}^{2}\left[ \int_{0}^{1}\frac{\mathrm{d}u}{|s^{2}u-|a||^{\frac{1}{4%
}}}\right] \mathrm{d}s\Big)
\end{eqnarray*}%
for a suitable constant $C<\infty $. Since%
\begin{equation*}
\int_{0}^{2}\frac{\mathrm{d}s}{|s^{2}-|a||^{\frac{1}{4}}}=\int_{0}^{2}\frac{%
\mathrm{d}s}{(s+|a|^{\frac{1}{2}})^{\frac{1}{4}}|s-|a|^{\frac{1}{2}}|^{\frac{%
1}{4}}}\leq \int_{0}^{2}\frac{\mathrm{d}s}{|s-|a|^{\frac{1}{2}}|^{\frac{1}{2}%
}}\leq 4 
\end{equation*}%
and%
\begin{equation*}
\int_{0}^{2}\int_{0}^{2}\frac{\mathrm{d}s\text{ }\mathrm{d}u}{|s^{2}u-|a||^{%
\frac{1}{4}}}=\int_{0}^{2}\frac{1}{s^{\frac{1}{2}}}\left( \int_{0}^{2}\frac{%
\mathrm{d}u}{|u-\frac{|a|}{s^{2}}|^{\frac{1}{4}}}\right) \mathrm{d}s\leq 8%
\text{ }, 
\end{equation*}%
we obtain that%
\begin{equation*}
\left\vert \mathrm{Re}\{\mathcal{K}\}\right\vert \leq 2^{4}C\left\Vert
f\right\Vert _{C^{1}}\left( 1+a^{2}+b^{2}\right) \text{ }. 
\end{equation*}

Similarly, 
\begin{eqnarray*}
\mathrm{Im}\{\mathcal{K}\} &=&-2\int_{0}^{r}\int_{-1}^{1}s^{d-1}\text{ }%
\widetilde{g}(s,u)\frac{bs^{d-1}\text{ }\widetilde{g}(s,u)}{%
(a-(2s)^{2}u)^{2}+b^{2}}\text{ }\mathrm{d}u\,\mathrm{d}s \\
&=&\frac{1}{2}\int_{0}^{r}s^{d-3}\left[ \int_{-1}^{1}\widetilde{g}%
(s,u)\left( \frac{\mathrm{d}}{\mathrm{d}u}\arctan \left[ \frac{a-(2s)^{2}u}{%
|b|}\right] \right) \mathrm{d}u\right] \mathrm{d}s \\
&=&\frac{1}{2}\int_{0}^{r}s^{d-3}\text{ }\widetilde{g}(s,1)\arctan \left[ 
\frac{a-(2s)^{2}}{|b|}\right] \mathrm{d}s \\
&&-\frac{1}{2}\int_{0}^{r}s^{d-3}\text{ }\widetilde{g}(s,-1)\arctan \left[ 
\frac{a+(2s)^{2}}{|b|}\right] \mathrm{d}s \\
&&-\frac{1}{2}\int_{0}^{r}s^{d-3}\left[ \int_{-1}^{1}(\partial _{u}%
\widetilde{g})(s,u)\arctan \left[ \frac{a-u(2s)^{2}}{|b|}\right] \mathrm{d}u%
\right] \mathrm{d}s
\end{eqnarray*}%
and $|\arctan (\gamma )|\leq \frac{\pi }{2}$ immediately implies that%
\begin{equation*}
\left\vert \mathrm{Im}\{\mathcal{K}\}\right\vert \leq C\left\Vert
f\right\Vert _{C^{1}} 
\end{equation*}%
for a suitable constant $C<\infty $.

\hfill $\square $

\bigskip \noindent \textbf{Acknowledgements: }This research is supported by
the FAPESP (grant 2016/02503-8), the CNPq and the Spanish Ministry of Economy and
Competitiveness MINECO: BCAM Severo Ochoa accreditation SEV-2013-0323 and
MTM2014-53850.

\end{document}